\documentclass[a4paper, amsfonts, amssymb, amsmath, preprint, 
showkeys, nofootinbib, twoside]{revtex4}
\usepackage[english]{babel}
\makeatletter\AtBeginDocument{\let\@elt\relax}\makeatother
\usepackage{ulem}
\usepackage[utf8]{inputenc}
\usepackage{multirow}
\usepackage{hhline}
\usepackage{tikz}
\usepackage{makecell}
\usepackage{comment}

\definecolor{tcA}{rgb}{0,0,0}
\usepackage{array}
\usepackage{bm}
\newcolumntype{P}[1]{>{\centering\arraybackslash}p{#1}}

\newcommand{\nslash}{\kern 0.2 em n\kern -0.50em /}
\newcommand{\kslash}{\kern 0.2 em k\kern -0.45em /}
\newcommand{\pslash}{\kern 0.2 em p\kern -0.50em /}
\newcommand{\Sslash}{\kern 0.2 em S\kern -0.50em /}
\newcommand{\Pslash}{\kern 0.2 em P\kern -0.50em /}
\newcommand{\Dslash}{\kern 0.2 em D\kern -0.65em /\kern 0.15em}

\newcommand{\xbj}{x_{\!\scriptscriptstyle B}}

\newcommand{\be}{\begin{equation}}
\newcommand{\ee}{\end{equation}}
\newcommand{\bea}{\begin{eqnarray}}
\newcommand{\eea}{\end{eqnarray}}

\def\kt{k_\perp}

\def\ptq{p_\perp}
\def\pt{P_{hT}}

\newcommand{\cdott}{{\mskip -1.5mu} \cdot {\mskip -1.5mu}}
\newcommand{\bq}{\begin{eqnarray}}
\newcommand{\eq}{\end{eqnarray}}

\def\kt{k_\perp}

\def\lsim{\mathrel{\rlap{\lower4pt\hbox{\hskip1pt$\approx$}}\raise1pt\hbox{$<$}}}
\def\gsim{\mathrel{\rlap{\lower4pt\hbox{\hskip1pt$\approx$}}\raise1pt\hbox{$>$}}}
\def\nostrocostruttino#1\over#2{\mathrel{\mathop{\kern 0pt \rlap
{\hbox{$#1$}}} \hbox{\kern-.135em $#2$}}}

\usepackage{soul}

\usepackage[titletoc]{appendix}

\usepackage[pdftex, pdftitle={Article}, pdfauthor={Author}]{hyperref} 
\begin{document}
\title{A TMD-based model for Hadronization off heavy nuclei}

 \author{F. A. Ceccopieri}
\author{R. Dupré}   
    \email[Correspondence email address: ]{raphael.dupre@ijclab.in2p3.fr }
    \affiliation{Université Paris-Saclay, CNRS, IJCLab, 91405, Orsay, France }

\date{\today} 

\begin{abstract}
Semi-inclusive deep inelastic scattering off nuclei is a unique process to study 
the parton propagation mechanism and its modification induced by the presence of the nuclear 
medium. It allows us to probe the medium properties, particularly the cold nuclear matter
transport coefficient, which can be directly linked to the nuclear gluon density. 
We present here a model for hadron production in deep inelastic lepton-nucleus scattering,
which takes into account the hadronic transverse momentum of final state particles via 
transverse-momentum dependent (TMD) parton distributions and fragmentation functions.
We implement parton energy loss and hadronic absorption with a geometrical model of
the nucleus. The model is compared with the nuclear SIDIS multiplicity ratios and 
transverse-momentum broadening data from the CLAS, HERMES, and EMC collaborations, 
aiming for a simultaneous description of these data sets. We obtain a good 
agreement over the various nuclear targets and the wide kinematical range of those 
experiments. We best describe the data with a transport coefficient 
$\hat{q} = 0.3$ GeV/fm$^2$, and we highlight the importance and the role of correlations 
in extracting this quantity.
\end{abstract}

\keywords{first keyword, second keyword, third keyword}
\maketitle

\section{Introduction}

The fragmentation of quarks into hadrons, \textit{i.e.}~hadronization, is a complex 
dynamical process of QCD. It is extensively studied in the vacuum and can be 
described by perturbative QCD and non-perturbative fragmentation 
functions~\cite{ParticleDataGroup:2022pthFF}. This process is modified in the
presence of a medium, such that studying the modification of 
the hadron spectra allows the extraction of certain medium properties. 
Conversely, a medium with known properties can be used for the study of the 
space-time development of the hadronization process~\cite{Accardi:2009qv}. Such 
studies require careful consideration of the longitudinal and transverse components 
of the hadron spectra simultaneously. Here, we will
study this problem using a model based on transverse momentum dependent (TMD) 
parton distribution functions (PDF) and fragmentation functions (FF). Recently, 
much progress has been made on TMD studies for proton, deuteron, and 
nuclei~\cite{Avakian:2016rst,Bacchetta:2016ccz,Alrashed:2021csd}. We present here 
a TMD model for hadron 
production in heavier nuclei and test it against the nuclear hadronization data sets 
from the EMC~\cite{EuropeanMuon:1991jmx}, 
HERMES~\cite{Airapetian:2007vu, HERMES:2009uge, HERMES:2011qjb} 
and CLAS~\cite{CLAS:2021jhm} collaborations.

The present study includes two main nuclear effects: induced energy loss and 
nuclear absorption. The former is related to elastic collisions of the propagating parton,
and the latter to the inelastic collision of the prehadron with nucleons.
The interface of those two mechanisms is the production length, $L_P$, 
\textsl{i.e.} the length necessary for the quark to turn into a color-singlet 
prehadron. Energy loss and nuclear absorption are complementary, but there is no 
consensus in the literature on their respective quantitative contributions. The 
analysis presented in this paper aims at building a model 
that accommodates both effects and can describe an extensive
data set with proper tuning of the production length $L_P$ and 
the quark transport coefficient, $\hat{q}$. The latter is the key parameter of
induced energy loss calculations, defined as the amount of 
transverse momentum the parton gains per unit of length of traversed material.
The hadron production in semi-inclusive deep-inelastic scattering (SIDIS) off 
nuclei is the best process for extracting those quantities.

Several groups have performed similar work in the past~\cite{Kopeliovich:2003py,
Accardi:2002tv,Accardi:2005hk,Guiot:2020vsf}, often with a focus on energy 
loss~\cite{Arleo:2003jz,Domdey:2008aq,Gao:2010mj,Song:2010zza,Song:2013sja,
Song:2014sja,Ru:2019qvz,Brooks:2020fmf}. Monte-Carlo 
event simulation is also sometimes used to approach the problem in a fully consistent 
way~\cite{Gallmeister:2007an,Ke:2023xeo} with the advantage of naturally 
including correlations. We aim to extend these previous works by using a TMD framework 
to describe the longitudinal and transverse components, as well as their correlations.

\section{TMD cross sections}
\noindent
We consider the process
\begin{equation}
  \label{sidis}
\ell(l) + N(P) \to \ell(l') + h(P_h) + X ,
\end{equation}
where $\ell$ denotes the beam lepton, $N$ the nucleon target, $h$ the produced hadron and 
$X$ the remainder of the hadronic final state. Particles four-momenta are given in parentheses. 
As customary, we define the photon momentum to be $q = l - l'$, and we introduce the invariants 
$Q^2 = - q^2$, $\nu = p \cdot q / M$, $\xbj = Q^2/ 2\,P \cdott q$, $y = P \cdott q/ P \cdott l$, 
$W^2=(P+q)^2$, and $z = P \cdott P_h / P\cdott q$. We define $P_{h \perp}$ to be 
the transverse components of hadron momentum $P_h$ with respect to the virtual photon 
momentum, defined in the target rest frame and $\phi_h$ the angle between the 
leptonic and the hadronic planes following the Trento convention~\cite{Bacchetta:2004jz}. 
Throughout this paper, we work in the one-photon exchange approximation and neglect the lepton 
mass. We retain in the calculation, when necessary, the nucleon mass $M$ 
and the hadron $h$ mass, $M_h$. The semi-inclusive lepton-hadron cross-section, summed over the helicities of the incoming lepton, can be written as~\cite{Bacchetta:2006tn}
\begin{eqnarray}
\frac{d^5\sigma}{d\xbj \, dy \,dz\, d P_{h\perp}^2 d\phi_h} 
&=& \frac{\pi \alpha^2}{Q^2 \xbj y }\, \Bigl\{
[1+(1-y)^2] F_{UU,T}+2(2-y)\sqrt{1-y} F_{UU,T}^{\cos \phi_h} \cos \phi_h + \nonumber\\
 && \hspace{1.5cm} +2(1-y) F_{UU,T}^{\cos (2\phi_h)} \cos (2\phi_h) \Bigr\}\,,
\label{e:crossmaster}
\end{eqnarray}
where the structure functions on the r.h.s. depends 
on $\xbj$, $Q^2$, $z$ and $P_{h\perp}^2$. The first and second subscripts of the above structure 
functions indicate the respective polarization of the beam and the target, whereas the third 
subscript specifies the polarization of the virtual photon. As all our analysis is performed 
at the lowest order in the strong coupling, the contributions corresponding to photon longitudinal polarization are absent at this order of accuracy. The structure function $F_{UU ,T}$ 
can be expressed as a convolution over partonic transverse momenta as
\begin{equation}
F_{UU ,T} = \xbj 
\sum_q e_q^2 \int d^2 \bm{p}_\perp\,  d^2 \bm{k}_\perp
\, \delta^{(2)}\bigl(\bm{p}_\perp + z\bm{k}_\perp^{} - \bm{P}_{h \perp} \bigr)
f_{q/P}(\xbj,Q^2,\bm{k}_\perp)\,D_{h/q}(z,Q^2,\bm{p}_\perp) ,
\label{e:convolition}
\end{equation}
where $f_{q/P}(\xbj,Q^2,\bm{k}_\perp)$ and $D_{h/q}(z,Q^2,\bm{p}_\perp)$ represent the TMD unpolarised parton distribution in the proton and TMD unpolarised fragmentation functions, respectively. Here $\bm{k}_\perp$ represents the transverse momentum of the initial state parton 
with respect to the virtual photon direction and $\bm{p}_\perp$ the transverse momentum of the detected hadron relative to the fragmenting parton. Both distributions are evaluated at the virtuality $Q^2$ of the photon.
The delta function results from transverse momentum conservation. The summation runs over quarks 
and antiquarks at this order. In the following, we will turn these expressions into a form suitable for phenomenological 
applications. Convolution integrals in Eq.~(\ref{e:convolition}) can be analytically performed 
under the hypothesis that the transverse momentum and the light-cone fraction dependencies of TMD 
densities could be factorized by assuming a Gaussian shape of the transverse part; such an 
assumption is known to be good at low transverse momentum. Therefore, we use 
for the TMD PDFs and FFs, the following ansatz:
\be
f_{q/P}(x_B,Q^2,\bm{k}_\perp) = f_{q/P}(x_B,Q^2) \, \frac{1}{\pi \langle\kt^2\rangle} \,
e^{-{\kt^2}/{\langle\kt^2\rangle}}
\label{partond}
\ee
and
\be
D_{h/q}(z,Q^2,\bm{p}_\perp) = D_{h/q}(z,Q^2) \, \frac{1}{\pi \langle p_\perp^2\rangle}
\, e^{-p_\perp^2/\langle p_\perp^2\rangle}.
\label{partonf}
\ee
Within this approximation, TMD PDFs and FFs are normalized to give the corresponding 
collinear distributions upon integration over transverse momentum.
In the present analysis, we include the evolution of the above collinear functions, as indicated by the explicit $Q^2$ dependence in the relevant functions, but we neglect the full TMD evolution, which would unnecessarily weight down the formalism and would require additional modelization in the nuclear case~\cite{Alrashed:2021csd}. This simplified approach also allows us to include the transverse momentum broadening 
when considering SIDIS on nuclear targets in a relatively economical way, which is 
crucial to us given the time-consuming cross sections evaluation of that specific case.

In such approximations, the structure function in Eq.~(\ref{e:convolition}) 
can be eventually rewritten as:
\begin{equation}
F_{UU,T}=\xbj \sum_q  e_q^2 f_{q/P}(\xbj,Q^2) \> D_{h/q}(z,Q^2)
\frac{e^{-\pt^2/\langle\pt^2\rangle}}{\pi\langle\pt^2\rangle}\,.
\label{FUUT}
\end{equation}
Its structure is dictated by the convolution over transverse momenta 
from Eq.~(\ref{e:convolition}) and the Gaussian assumption on the transverse part in Eq.~(\ref{partond}) and Eq.~(\ref{partonf}).
The full width $\langle\pt^2\rangle$
being the average hadronic transverse momentum is then of the form:
\be
\langle \pt^2 \rangle = \langle \ptq^2 \rangle + z^2 \langle \kt^2 \rangle\,.
\label{eq:meanpt}
\ee

In general, we do not expect  
the shape of the hadronic width $\langle \pt^2 \rangle$ to have a significant impact 
on our nuclear observables, which by construction isolate nuclear effects. However, 
some dependence may arise due to correlations in the integration of the model over the 
experimental bins, which will likely introduce some sensitivity on the $z$-shape of 
$\langle \pt^2 \rangle$. For that reason, we made a dedicated extraction of 
$\langle \pt^2 \rangle$ whose details are collected in Appendix \ref{appx}.

Finally, realistic results are obtained only after the proper implementation of the kinematic cuts
used in the data set we try to reproduce. 
In particular, the contribution of target fragmentation is not accounted for in our formalism. 
Therefore, we apply a cut on Feynman $x$, with $x_F=2 h_z^*/W$, $h_z^*$ being the hadron 
third component defined in the $\gamma^* P$ center-of-mass frame, whenever specified by the 
kinematical selection of a given data set. Next, we consider the cut on the invariant mass of 
the system $X$ ($M_X$), where the latter represents the hadronic final state but with the 
measured final state hadron $h$ excluded. In all our calculations, we enforce the condition 
$M_X>M$, which induces a sizeable decrease of the $P_{h\perp}$-width at large $z$ and low $\nu$. 

\section{Nuclear effects}

The two main observables used by the experiments are the multiplicity ratios
\begin{equation}
    R_h^A(\nu,Q^2,z,\pt^2)=\frac{N_h^A(\nu,Q^2,z,\pt^2)/N_e^A(\nu,Q^2)}{N_h^D(\nu,Q^2,z,\pt^2)/N_e^D(\nu,Q^2)},
\label{obs:multiplicity}    
\end{equation}
and the transverse momentum broadening
\begin{equation}
    \Delta \langle \pt^2 \rangle = \langle \pt^2 \rangle_A - \langle \pt^2 \rangle_D,
\label{obs:ptbroad}
\end{equation}
where $A$ stands for the atomic mass number, $N_e$ and $N_h$ the number of inclusive and 
semi-inclusive events, respectively, and the observables being both normalized to deuterium, $D$.
In the present analysis, we will not provide an extensive comparison of all particle species, 
but we will compare our model with multidimensional data sets whenever possible to account
for possible correlations in the data. For this reason, we focus on positively charged pions: 
they are abundantly produced, such that precise fragmentation functions are available. 
In addition, they are nearly massless, which avoids theoretical issues due to hadron 
mass corrections.

\subsection{Production length and nuclear medium density}

We implement the nuclear effects under the hypothesis that energy 
loss occurs before absorption. Thus, we need to know the length of the total traversed material
and the production length $L_P$, where parton energy loss ends and absorption starts. 
To go further and account for the density profile of the nucleus, we need to define effective
lengths. To do so, we assume that the incident photon scatters on a parton belonging to a nucleon 
at position $x_i,\vec{b}_i$, with the origin fixed to the nuclear center. The scattered parton 
then moves along the positive $x$ direction. We describe the nuclear matter distribution with
a Wood Saxon distribution, $\rho(x,\vec{b}_i)$, normalized to $A$. With this in mind, we adopt 
the following definition of total path length traversed in nuclear matter:  
\begin{equation}
L^*_T(x_i,\vec{b}_i) = 
\frac{1}{\rho(0,0)}
\int_{x_i}^{\infty} dx \rho(x,\vec{b}_i).
\label{eq:pathlength}
\end{equation}

The production length, \textsl{i.e.} the length traveled by the colored parton before forming 
a pre-hadron, is parametrized with the following form:
\begin{equation}
L_P(z,\nu)=N z^\lambda (1-z)^\beta \Big(\frac{\nu}{ \mbox{GeV}} \Big)^\gamma.
\label{eq:production_length}
\end{equation}
Generally, one expects $\lambda \simeq \beta \simeq \gamma \simeq 1$ to reflect 
the Lorentz boost and fragmentation constraints on the parton lifetime. These 
settings have not provided a satisfactory description of data in the past and have 
been modified or complemented in various ways in the literature~\cite{Accardi:2002tv,Arleo:2003jz,Kopeliovich:2003py,Accardi:2009qv}. 
Here, we introduce these parameters, but we adjust them to better describe the data. 
To limit the number of parameters, we do not explore the interesting possibility 
that $L_P$ might depend on other variables, such as $\pt^2$ or 
$Q^2$~\cite{Kopeliovich:2003py,Kopeliovich:2006xy}. Similarly to the total path length 
$L^*_T$, we define an effective  production length 
\begin{equation}
L_P^*(z,\nu;x_i,\vec{b}_i)= 
\frac{1}{\rho(0,0)}
\int_{x_i}^{x_i+ L_P(z,\nu)} dx \rho(x,\vec{b}_i)\,,
\label{eq:pathlength_eff}
\end{equation}
which, at variance with Eq.~(\ref{eq:pathlength}),
takes into account the finite production length $L_p$ in the upper integration limit.

\subsection{Partonic energy loss}

When partons traverse a QCD medium such as the nucleus, we expect them to emit gluons and 
lose some of their energy. Several groups have tackled the problem of calculating this energy 
loss with a rather large spread of results~\cite{Armesto:2011ht}. We chose to use the calculation
by Arleo~\cite{Arleo:2002kh}, based on the BDMPS calculations~\cite{Baier:1996kr,Baier:2001yt}. In 
this framework, the energy loss $\epsilon$ depends on the transport coefficient and the length of 
the medium through the characteristic gluon energy
\begin{equation}
w_c = \frac{1}{2} \, \hat{q} \, L^{*2}_p.
\label{eq:wc}
\end{equation}
The energy loss probability distribution $W(\epsilon)$ by Arleo includes also the dependence on the 
outgoing parton energy, which we take to be $\nu$. It is expressed as follows:
\begin{equation}
\overline{W} (\bar{\epsilon}, \bar{\nu}) = \frac{1}{\sqrt{2\,\pi}\, \sigma(\bar{\nu})\,\bar{\epsilon}}\,\exp\left[-\frac{\left(\log{\bar{\epsilon}}-\mu(\bar{\nu})\right)^2}{2\,\sigma(\bar{\nu})^2}\right] \, ,
\label{eq:qw}
\end{equation}
where barred variables are normalized to $\omega_c$, the functions $\mu$ and $\sigma$ are 
empirical~\cite{Arleo:2002kh} and its normalized version is defined by 
$\overline{W} (\bar{\epsilon} = \epsilon/\omega_c) = \omega_c W(\epsilon)$.
We model nuclear-modified FFs according to the kinematic rescaling proposed by Wang, Huang 
and Sarcevic~\cite{Wang:1996yh}, which amounts to shift the argument of the collinear FFs:
\begin{equation}
    z_* = \frac{z}{1- \epsilon /\nu } \, .
\label{eq:z-rescaling}    
\end{equation}
The implementation of such a rescaling allows us to obtain the nuclear-modified TMD FFs:
\begin{equation}
    z D_{h/q}^{A}(z, \nu, Q^2, p_\perp) = 
    \int_0 ^{(1-z)\nu}  d \epsilon \; W(\epsilon,\nu,\hat{q}, L_P^*)  z_*  
    D_{h/q}(z_*, Q^2,p_\perp)\,,
\label{eq:nFFs}    
\end{equation} 
which  generalizes the collinear variables rescaling proposed in  Ref.~\cite{Wang:1996yh} 
to transverse ones. For clarity, we have omitted 
its dependence via $L_P^*$ on the position $x_i,\vec{b}_i$
of the nucleon in the nucleus on which the incident photon scatters.
Moreover, we use the connection proposed by BDMPS to link the energy loss 
$\epsilon$ to $\langle q_\perp^2 \rangle$, the average transverse momentum acquired by 
the parton traversing the medium
\cite{Baier:1996sk,Baier:1998kq}:
\begin{equation}
-\frac{dE}{dx} = \frac{\alpha_s N_C}{4} \langle q_\perp^2 \rangle, 
\label{eq:BDMPS}
\end{equation}
with $E$ the energy of the gluon emitter, $x$ is the distance traversed by the parton, 
$N_C$ the number of colors, $\alpha_s$ the strong coupling constant.
From this, we can link the transverse momentum broadening to the parton energy loss:
\begin{equation}
\langle q^2_\perp \rangle = \frac{4 \epsilon}{ \alpha_s N_C L_P^*},
\label{eq:BDMPS2}
\end{equation}
We assume that this source of transverse momentum is distributed according to a Gaussian 
of width $\langle q^2_\perp \rangle$.
Since it is acquired during the in-medium fragmentation process, its
width is scaled by a factor $z_*^2$. As a result, the overall nuclear-modified fragmentation 
width is given by  
$\langle p^2_\perp \rangle_A = \langle p^2_\perp \rangle (z_*) + z_*^2 \langle q^2_\perp \rangle$. 
Within these approximations, the nuclear-modified fragmentation function in Eq.~(\ref{eq:nFFs}) 
can be written as
\begin{equation}
D_{h/q}^{A}(z,\nu, Q^2, p_\perp) = 
    \int_0 ^{(1-z)\nu}
    d \epsilon \frac{W(\epsilon,\nu,\hat{q}, L_P^*)}{1-\epsilon/\nu} D_{h/q}(z_*,Q^2) \frac{e^{-p_\perp^2/\langle p^2_\perp \rangle_A}}{\pi\langle p^2_\perp \rangle_A}\,.
\end{equation}

\subsection{Absorption} 

After the production length, a prehadron starts to form and might interact with the medium, thus
not giving rise to a measured hadron in the final state. To simulate this, we use an attenuation 
factor of the form: 
\begin{equation}
    n_{att}(z,\nu;x_i,b_i) = 
    1 - \Bigg[\frac{ L^*_T-L_P^*}{L_{typical}}\Bigg]^d,
    \label{eq:absorption}
\end{equation}
with $L_{typical}$ and $d$ free parameters. This parametrization is constructed 
such as to introduce a correlation between 
the length traversed by the prehadron in the medium, $L^*_T- L_P^*$, and the absorption process.
Given the definition in Eq.~(\ref{eq:pathlength}) and Eq.~(\ref{eq:pathlength_eff}), 
the latter quantity is greater than zero by construction and, at any fixed value of $L^*_T$, 
the attenuation is maximal when  $L^*_p \rightarrow 0$.
The attenuation factor $n_{att}$ implicitly depends on $z$ and 
$\nu$ through the dependence of $L_P$ on those variables. We further assume that it is quark-flavor
independent and apply it as a multiplicative factor in defining nuclear FFs in Eq.~(\ref{eq:nFFs}).

We do not attribute transverse momentum in association with the pre-hadron interaction as the elastic
process contributes very little compared to inelastic processes~\cite{Domdey:2008aq}. Thus, we
consider here that the pre-hadrons are absorbed in the nucleus, giving rise to fragments in the 
target fragmentation region outside our model's domain.

\subsection{Fermi motion}

Among the initial state effects we need to consider is the Fermi motion
of the nucleons in the nuclear targets. We use the parametrizations of 3-momentum distributions, 
$n_A(\vec{P})$ presented in Ref.~\cite{CiofidegliAtti:1995qe}, with 
$\vec{P}$ the nucleon three-momentum in the nucleus.
We neglect the effect 
of Fermi motion in the longitudinal direction due to its small size relative to the typical energies 
involved in the experiments we aim to describe. As we only consider the transverse contribution,
we define the average transverse Fermi motion as
\begin{equation}
\langle P_{T}^2 \rangle_A = \frac{ \int d^3 \vec{P} \; P_{T}^2 \; n_A(\vec{P}) }{ \int d^3 \vec{P} \;  n_A(\vec{P})}, 
\end{equation}
where adopting spherical coordinates, the transverse momentum is defined with respect to the 
$z$-axis, corresponding to the incoming photon direction. Projecting out transverse components, 
the integral can be written as 
\begin{equation}
\langle P_{T}^2 \rangle_A = \frac{2}{3} \int_{0}^{\infty} \; P^4 \; dP \; n_A(P)\,. 
\end{equation}
The calculation returns an average transverse Fermi motion of around 0.011 Ge$\mbox{V}^2$ for 
deuterium which rises fast to 0.036 Ge$\mbox{V}^2$ for carbon, and 
then flattens out for heavier nuclei and ends around 0.044 Ge$\mbox{V}^2$ for a lead nucleus.

This source of initial state transverse momentum, assuming it is distributed as a Gaussian with widths provided by the calculation, adds to the initial state partonic transverse momentum. 
Therefore, as dictated by transverse kinematics in eq.~(\ref{eq:meanpt}), the corresponding widths will sum up and will be weighted by a factor $z^2$.

\subsection{Nuclear structure functions}

Putting together all the elements discussed above, the unpolarized nuclear structure function reads
\begin{multline}
F_{UU,T}^A=\xbj \sum_q  e_q^2 f_{q/A}(\xbj,Q^2) 
\int dx_i \int d^2 \vec{b}_i 
\frac{\rho(x_i,b_i)}{A} 
    \int_0 ^{(1-z)\nu}
    d \epsilon \frac{W(\epsilon,\nu,\hat{q}, L_P^*)}{1-\epsilon/\nu} \\
    n_{att}(z,\nu;x_i,b_i)
     D_{h/q}(z_*,Q^2) \frac{e^{-\pt^2/\langle\pt^2\rangle(z_*)}}{\pi\langle\pt^2\rangle(z_*)}\,,
\end{multline}
where $f_{q/A}$ are the nuclear PDFs
for which we adopt the set \texttt{nCTEQ15}~\cite{Kovarik:2015cma}, and
$D_{h/q}$ are the vacuum parton-to-pion fragmentation functions 
for which we adopt the \texttt{DEHSS} set~\cite{deFlorian:2014xna}.
In general, for the observable studied in the present paper,  
the predictions are rather independent of the choices of the specific sets of PDFs and FFs.

We assume that all sources of transverse momentum broadening have a Gaussian distribution 
such that their convolution gives an overall Gaussian with a width 
\begin{equation}
\langle \pt^2 \rangle_A = 
 \langle p^2_\perp \rangle (z_*) + z^2 [ \langle k^2_\perp \rangle + \langle P_{T}^2 \rangle_A]
+ z_*^2 \langle q^2_\perp \rangle.
\label{nuclear_pt}
\end{equation}
The first term is the width of the fragmentation function but evaluated, according to 
Eq.~(\ref{eq:nFFs}), at $z_*^2$. The terms weighted by $z^2$ are for the initial state 
parton momenta intrinsic in the nucleon and the nuclear Fermi motion. The last term 
accounts for the nuclear broadening due to parton energy loss and is weighted by $z_*^2$.

The observables introduced in Eq.~(\ref{obs:multiplicity}) and Eq.(~\ref{obs:ptbroad}) are both 
normalized to the deuterium. Such a normalization point is commonly used since deuterium does not show any nuclear effect, allowing hadronization to be described in terms of vacuum fragmentation functions. 
However, the model discussed so far, designed to describe hadronization off heavy nuclei, shows sizeable nuclear effects when extrapolated to very light nuclei like deuterium. Therefore, to ensure a proper subtraction of spurious effects and a consequent correct normalization of the observables, we normalize the predictions to our extrapolated version of deuterium.

\section{Results}

The large number of convolutions makes theory predictions time-consuming and not suitable for a fit. To overcome this problem, we first studied the model's behavior in parameter space and then fixed the latter via a tuning procedure that allows us to fairly describe all data sets at once.

\subsection{Energy loss only parameterization}

Among the two mechanisms we consider, the partonic energy loss has been extensively studied 
in the literature and has a solid theoretical background. Therefore, in the first attempt, we tried to describe the data with that mechanism alone responsible for both the depletion of hadron yields and the broadening of the transverse momentum. 
\begin{figure*}
\begin{center}
\includegraphics[scale=0.6]{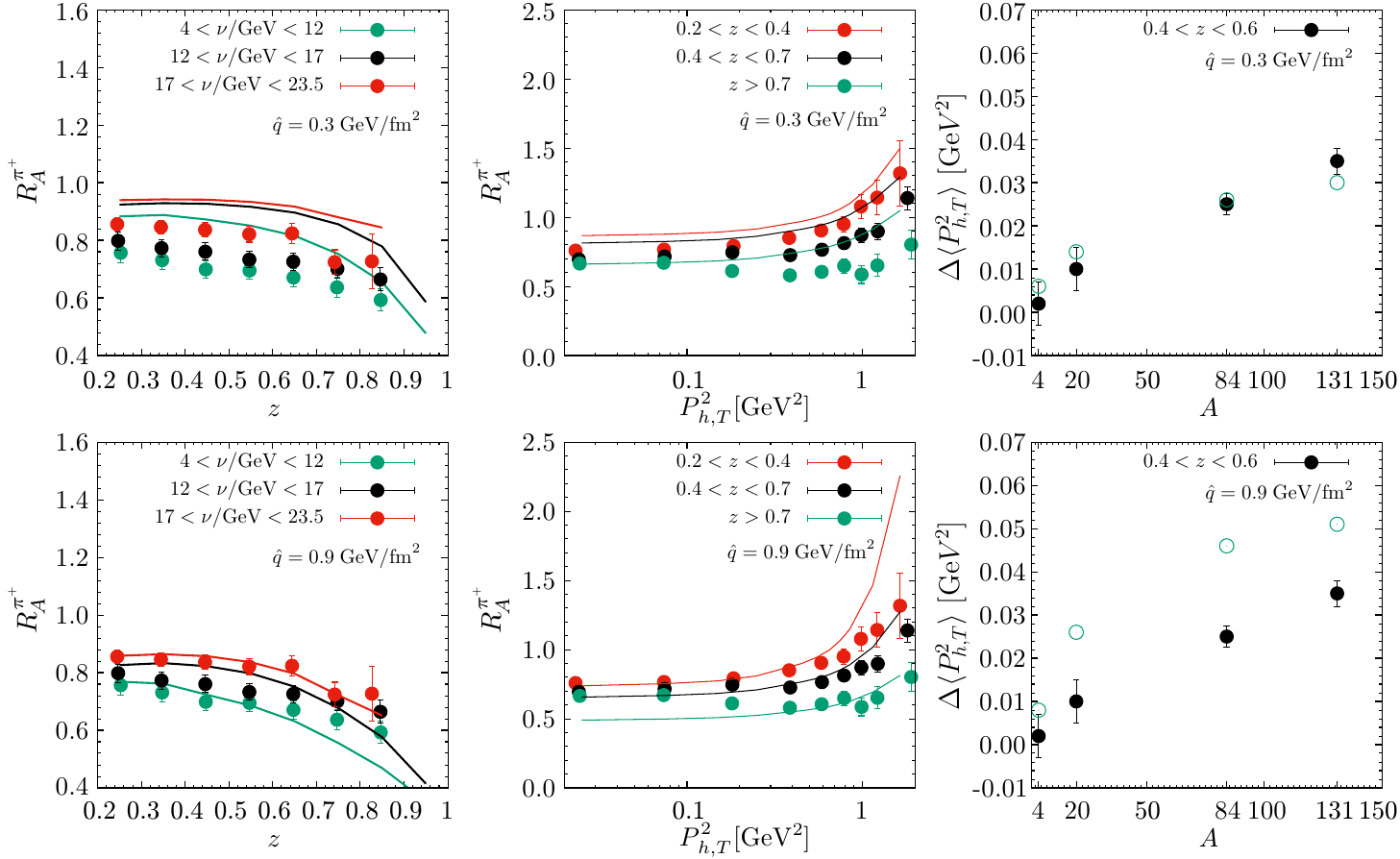}
\caption{Predictions within an ``energy loss alone'' model, \textsl{i.e.}  $L_P \rightarrow \infty$, with 
$\hat{q}=0.3$ GeV/f$\mbox{m}^2$ in top panels 
and $\hat{q}=0.9$ GeV/f$\mbox{m}^2$ in bottom ones.
We present the $\pi^+$-multiplicity ratios on a Krypton target as a function of $z$ in the leftmost column 
and as a function of $P_{hT}^2$ in the middle one. 
In the rightmost column, we present the $\Delta \langle \pt^2 \rangle$-broadening as a function of $A$.
Data from the HERMES collaboration~\cite{HERMES:2009uge}.}
\label{plot:Hermes_Kr_Rmulti_noabs}
\end{center}
\end{figure*}

Within this picture, the parton will radiate all the way out of the nucleus, \textsl{i.e.} $L_P \rightarrow \infty$, maximizing the amount of radiated energy, the hadronization will occur outside the nucleus, and no absorption will take place, \textsl{i.e.} $n_{att}=1$.
Under this approximation, we present in the top panels of Fig.~(\ref{plot:Hermes_Kr_Rmulti_noabs})
the comparison of the HERMES multiplicity ratios as a function of $z$ and 
$\pt^2$ with our default value  
$\hat{q}$ = 0.3 GeV/fm$^2$. 
The  multiplicity ratios show a clear deficit in normalization (leftmost panels) 
while the transverse momentum broadening, $ \Delta \langle \pt^2 \rangle$, is fairly reproduced (rightmost panel). 
The normalization issue of the former can be alleviated by using a larger value of $\hat{q} \approx 0.9$~GeV/fm$^2$, 
as shown in the bottom left panels of Fig.~(\ref{plot:Hermes_Kr_Rmulti_noabs}). 
However, such a large $\hat{q}$ entering in Eq.~\ref{eq:BDMPS}, leads to an abnormally large $ \Delta \langle \pt^2 \rangle$, by far greater than the one seen in the data (bottom right panel).
We conclude that with an energy loss model alone, we can not describe those 
observables simultaneously with a single value of $\hat{q}$.
However, such an issue can be overcome by assuming 
that another mechanism is at work to reduce the yields, \textsl{i.e.} absorption. The attenuation   
is produced by inelastic interactions of the so-called prehadrons, a colorless state yet to become 
real hadron,  with the medium, possibly creating secondary particles with lower energies. 
Under the hypothesis that elastic prehadron-nucleon cross sections are small, this mechanism generates neither broadening nor energy loss.

\subsection{Full model parametrization}

We include the {\it ad hoc} absorption from Eq.~\ref{eq:absorption} for which 
we need to find suitable parameters. Our aim is not to make the best possible description of 
the data in all corners like in a fit. To adjust our parameters, we concentrate on kinematics 
which are the safest in terms of TMD factorization~\cite{Boglione:2022gpv} ($0.3 < z < 0.8$), 
parton energy loss applicability ($\nu > 10$), and the Gaussian TMD model 
($P_{h\perp}^2 < 0.5$~GeV$^2$). We will check later whether our model can 
be extrapolated out of this kinematic region, so we concentrate here on a subset of the HERMES 
data~\cite{HERMES:2009uge,HERMES:2011qjb} to tune our parameters.

\begin{figure*}
\begin{center}
\includegraphics[scale=0.5]{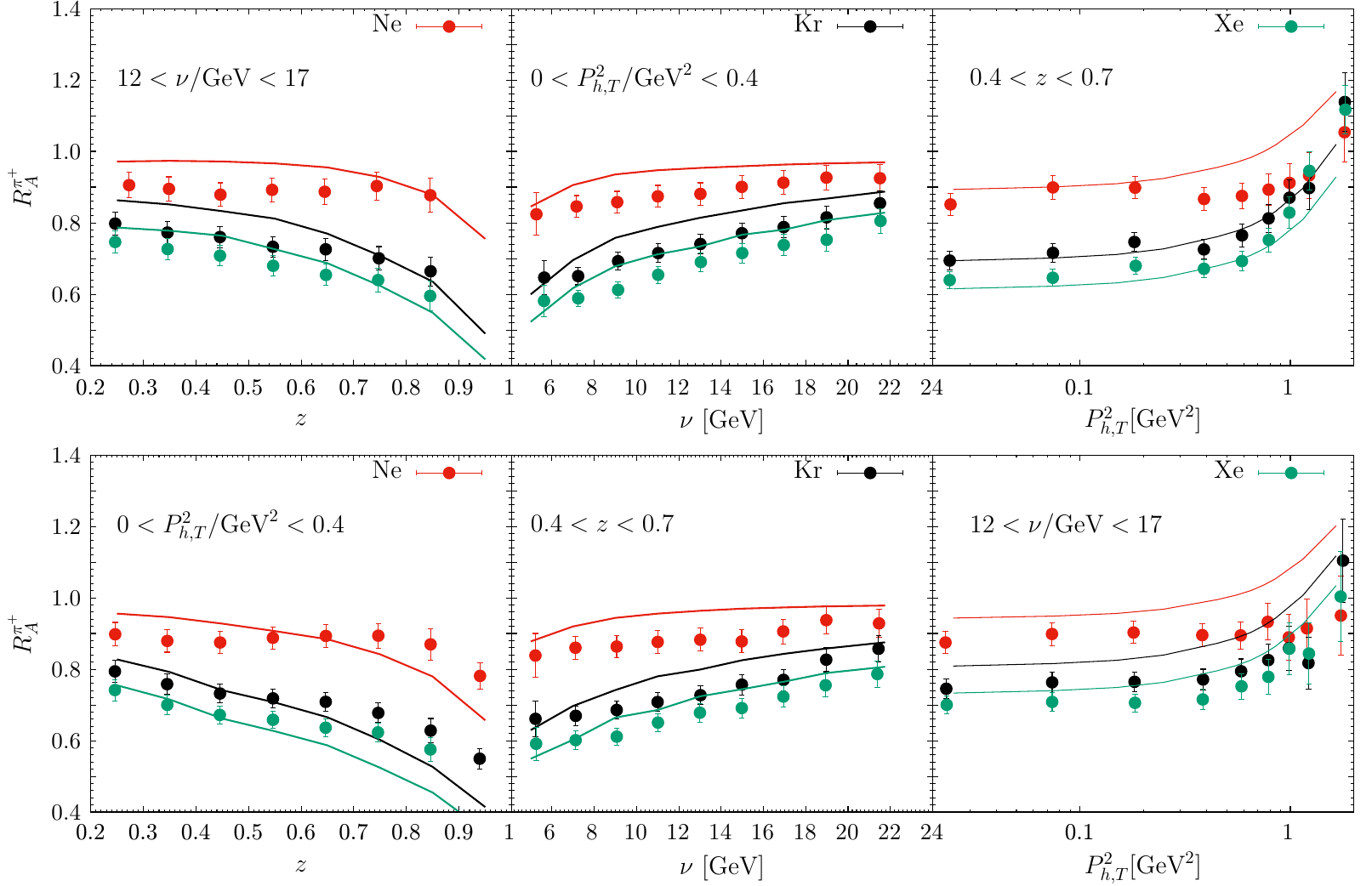}
\caption{ Multiplicity ratio for $\pi^+$ on different nuclear targets as a function of various kinematical variables in selected ranges compared to our model with parameters as in Tab.~[\ref{tab:params}]. Data from the HERMES collaboration~\cite{HERMES:2009uge}.}
\label{plot:Hermes_Rmulti_vs_A}
\end{center}
\end{figure*}

\begin{figure}
\begin{center}
\includegraphics[scale=0.6]{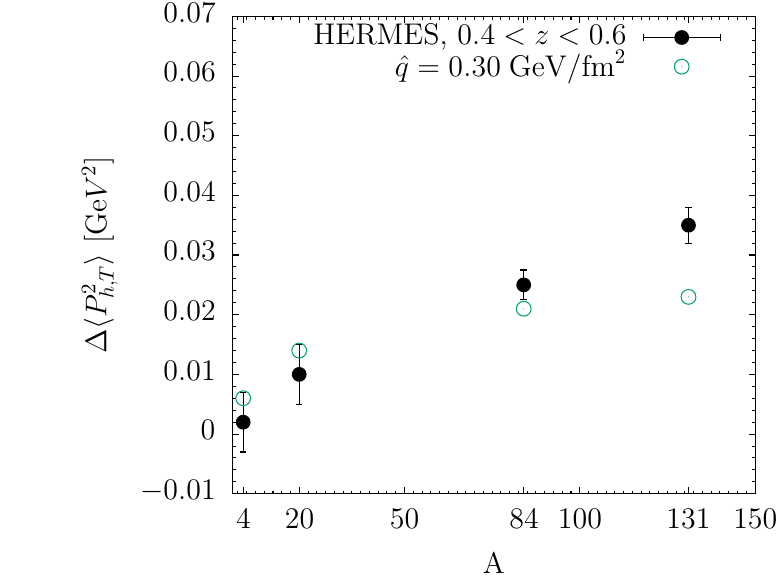}
\caption{The $\pi^+$ transverse momentum broadening in the range $0.4<z<0.6$ as a function of $A$ compared to our model with parameters as in Tab.~[\ref{tab:params}]. Data from the HERMES collaboration~\cite{HERMES:2009uge}.}
\label{plot:Hermes_ptbroad_vs_A}
\end{center}
\end{figure}

We show in Fig.~\ref{plot:Hermes_Rmulti_vs_A} and Fig.~\ref{plot:Hermes_ptbroad_vs_A} 
a selection of data according to the criteria listed above together with the model with 
adjusted parameters summarized in Tab.~\ref{tab:params}. The overall agreement is
satisfactory, but a few preliminary comments can be made immediately about the parameters 
obtained. We notice that the exponent parameters for the production length ($\lambda$, 
$\beta$ and $\gamma$) are small compared to previous modelisations~\cite{Accardi:2009qv}. 
In particular, with $\lambda=0$ and $\beta=0.25$, we significantly suppress the $z$ dependence 
of $L_P$ to describe the data adequately. This is needed so as not to over-suppress the hadron 
yield at both $z$ extremes. We show in Tab.~\ref{tab:LP} the typical production lengths 
obtained with this set of parameters for the CLAS, HERMES, and EMC data. We observe that 
these results are consistent with previous work~\cite{Accardi:2009qv}.
 
\begin{table}
\setlength\extrarowheight{2pt}
\begin{center}
\begin{tabular}{ P{1.875cm} P{1.875cm} P{1.875cm} P{1.875cm} } 
\hline \hline 
 \multicolumn{4}{c}{ Production length} \\
\hline 
   $N$       &  $\lambda$ &  $\beta$ & $\gamma$  \\ 
   4.0 fm    &     0.0    &    0.25  &    0.3     \\ 
\end{tabular}

\begin{tabular}{ P{2.3cm} P{1.3cm} P{0.5cm} P{1.7cm} P{1.7cm} } 
\hline \hline
 \multicolumn{2}{c}{Transport coefficient} & &  \multicolumn{2}{c}{Nuclear absorption} \\
\hline
  $\hat{q}$ & $\alpha_s$ & & $L_{typical}$          & $d$     \\
  0.3 GeV/fm$^2$  & 0.5      & & 2 \mbox{fm}           & 0.5   \\
\hline 
\end{tabular}
\end{center}
\caption{Numerical values of the parameters controlling the production length, the transport coefficient and nuclear absorption.} 
\label{tab:params}
\end{table}

\begin{table}
\setlength\extrarowheight{2pt}
\begin{center}
\begin{tabular}{ P{1.5cm} P{1.5cm} P{1.5cm} P{1.5cm} P{1.5cm} } 
\hline \hline
       &    $E_{beam}$   &    $\langle \nu \rangle$    &    $\langle z \rangle$  &    $L_P$  \\
       &    [GeV]        &   [GeV] &     &    [fm]\\
\hline
CLAS   &    5                &      3          &      0.5         &     4.7  \\
HERMES &    23.5             &      10         &      0.5         &     6.7  \\
EMC    &    280              &      100        &      0.35        &     14.3  \\
\hline 
\end{tabular}
\end{center}
\caption{Values of $L_P$ in the average kinematics of different experiments whose data are considered in this paper.}  
\label{tab:LP}
\end{table}

We obtain a transport coefficient value for cold nuclear matter, $\hat{q} = 0.3$ GeV/fm$^2$,
which is larger than some of the most recent extractions~\cite{Ru:2019qvz,Brooks:2020fmf}. 
However, these previous works neglect the correlation between longitudinal and transverse 
components. We attribute our different result to a survivor bias: because the measured 
transverse momentum broadening is obtained only through the detected hadrons, it is thus 
indirectly affected by absorption. In the HERMES data, the hadron yield is reduced by 
20-30\% in Krypton, whatever model is used, it is clear that the suppressed part of 
the spectrum will be strongly correlated to the length of nuclear material traversed. So 
effectively, the transverse momentum observables select hadrons that have interacted less 
with the nuclear medium. Thus, it creates a correlation between absorption and energy loss 
that reduces the measured transverse momentum broadening and motivates the use of a larger 
$\hat{q}$. This effect can already be observed in our first test with energy 
loss only in Fig.~\ref{plot:Hermes_Kr_Rmulti_noabs}. There, it is clear that despite a 
factor 3 in $\hat{q}$ between the two scenarios explored, the transverse momentum broadening 
increases by less than a factor 2 for krypton. Together, our results indicate that this 
effect is large no matter what the source of the hadron yield quenching is, and thus it 
needs to be accounted for in the description of the transverse-momentum dependent observables.

With the hadron absorption parameters, we can notice that the
coefficient $d$ indicates the dependence on the path length. Our best value, $d=0.5$, 
indicates that the pre-hadron absorption is reducing as it propagates. Such behavior is possible
through a filtering effect where compact prehadrons survive longer, and thus, the absorption 
effect drops. However, this is in direct contradiction with the results of previous studies
~\cite{Kopeliovich:2003py,Guiot:2020vsf,Gallmeister:2007an}, which would favor $d=1$ or $2$. 
The differences between the results presented in these studies and our approach are 
numerous, but the key difference is likely linked to the way $L_P$ is parameterized.
Finally, on the question of hadron absorption, we note that the value of $L_{typical}$ is 
rather small and hints at a rather large prehadron-nucleon cross-section.

\subsection{Comparison to HERMES data}

\begin{figure*}
\begin{center}
\includegraphics[scale=0.5]{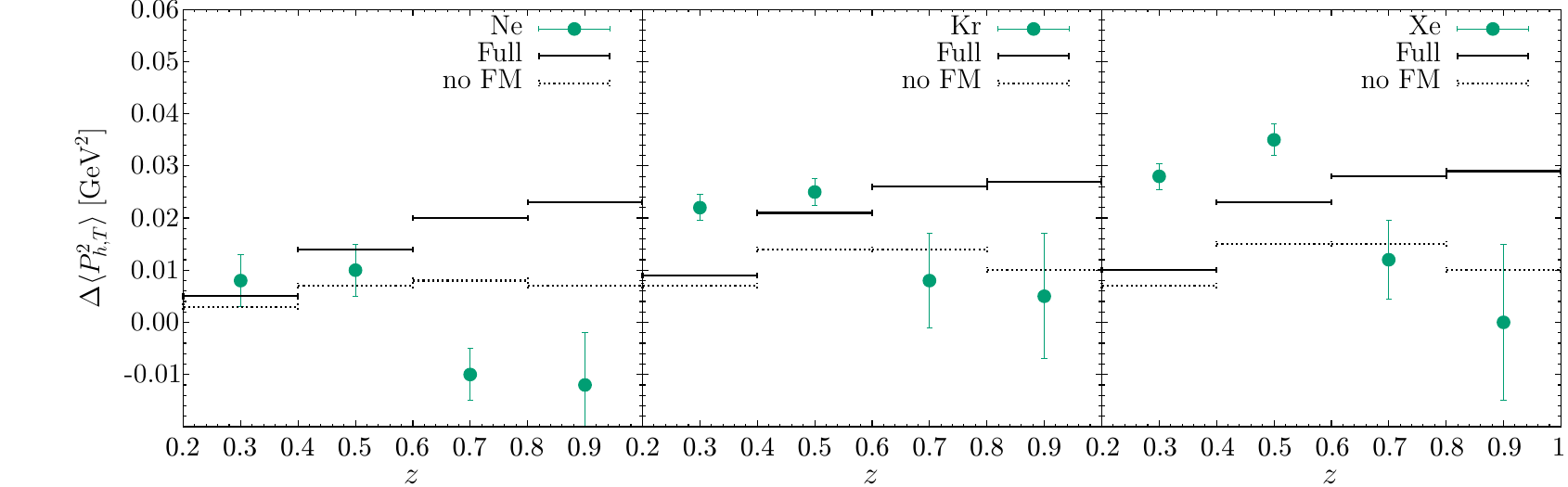}
\caption{Transverse momentum broadening on various heavy nuclei as a function of $z$ compared to HERMES data of Ref. \cite{HERMES:2009uge}. Our model, with parameters as in Tab.~[\ref{tab:params}], is shown with (solid lines) and without (dotted) Fermi motion contribution.}
\label{plot:Hermes_avept2}
\end{center}
\end{figure*}

Going into more details on Fig.~\ref{plot:Hermes_ptbroad_vs_A}, we observe that the $A$ 
dependence of $\Delta p_T^2$ is not completely satisfactory, as the model overshoots the 
light nuclei and undershoots heavier nuclei. This problem is a common feature of numerous 
descriptions of this observable in the literature~\cite{Domdey:2008aq,Ru:2019qvz,
Song:2014sja,Brooks:2020fmf}. The best description we obtain of this observable is in
Fig.~\ref{plot:Hermes_Kr_Rmulti_noabs}, where production length is set to infinity and 
absorption is thus completely deactivated. It appears impossible to resolve this issue with
existing model frameworks because reducing the production length to increase the 
absorption contribution further degrades the description of this observable. We 
conclude that all the models (ours included) are still missing an element contributing 
to the shape of the transverse momentum broadening as a function of the atomic number $A$.
\begin{figure*}
\begin{center}
\includegraphics[scale=0.5]{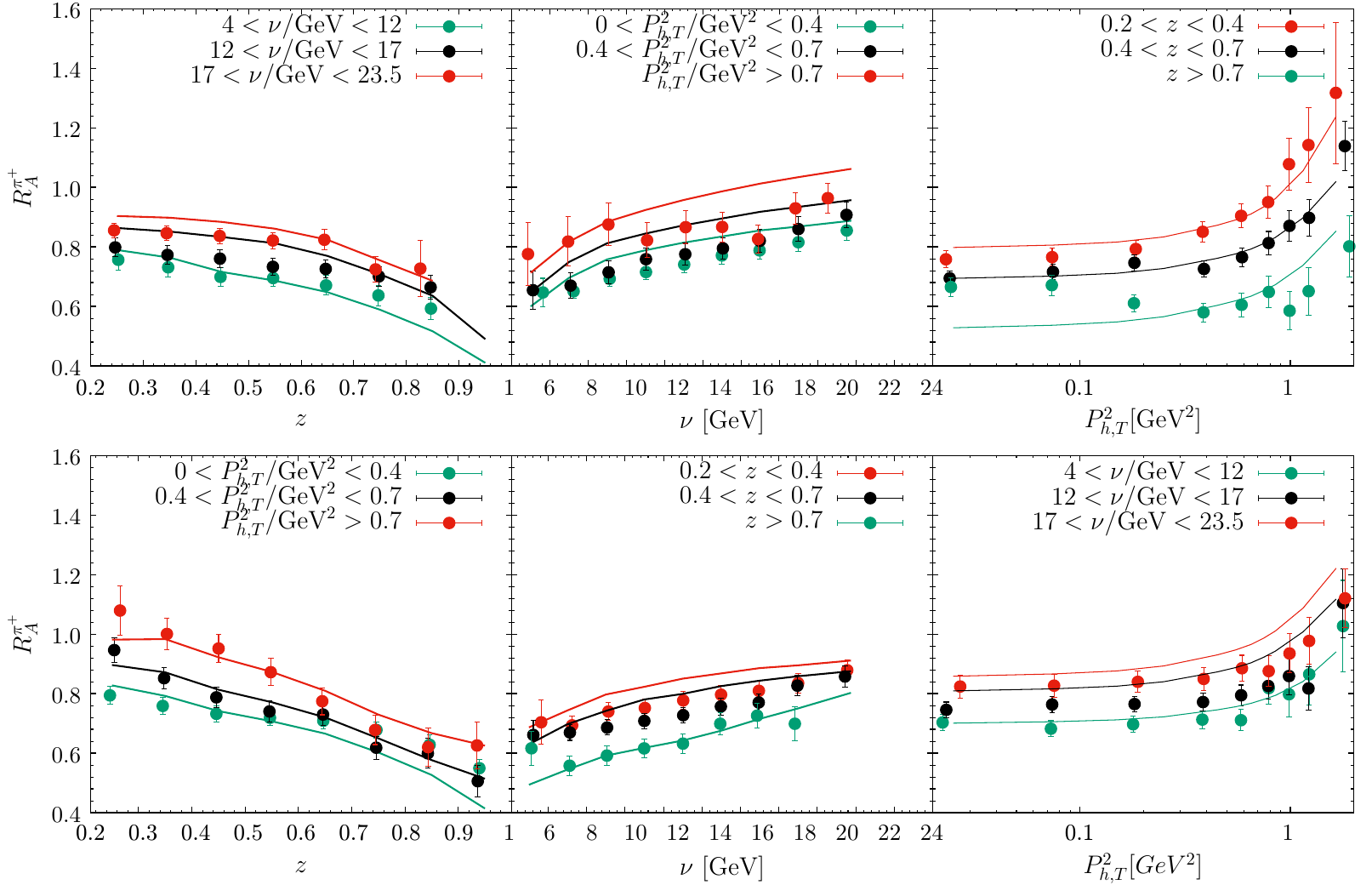}
\caption{Multiplicity ratio for $\pi^+$ on a Krypton target as a function of kinematic variables compared to our model with parameters as in Tab.~[\ref{tab:params}]. Data from the HERMES collaboration~\cite{HERMES:2009uge}.}
\label{plot:Hermes_Kr_Rmulti}
\end{center}
\end{figure*} 
We compare our model to more HERMES multiplicity ratio data in Fig.~\ref{plot:Hermes_Kr_Rmulti},
and note that all displayed dependencies are pretty well reproduced. Some of the high 
$z$ behaviors are missed. However, as discussed above, this is a phase space where TMDs 
do not dominate the cross-section. Thus, we do not expect our model to describe this phase
properly. To further analyze the behavior of the model, 
we compare predictions for transverse momentum broadening
on different targets as a function of $z$ in Fig.~(\ref{plot:Hermes_avept2}). We observe that the transverse momentum broadening shape as 
a function of $z$ appears to be largely off for all nuclear targets. At low $z$, this is 
to be expected as target fragmentation might contribute, but at large $z$ the discrepancy is more 
puzzling. Given the structure of Eq.~(\ref{nuclear_pt}), the rising pattern 
of $\Delta \langle \pt^2 \rangle$ against $z$ is however expected. Of the previously cited 
literature, only Domdey et al.~\cite{Domdey:2008aq} shows the same observable, reporting 
a similar discrepancy. Furthermore, we show our model
with and without Fermi motion in the same figure. We observe that the latter is the main contribution at 
large $z$. The absence of such a basic effect in the data is surprising. We suspect correlated errors 
in the data to be at the origin of this behavior, as all nuclei are normalized to the same 
deuterium data. Otherwise, that would indicate that factorization is largely broken in this 
kinematical region. To reduce the Fermi motion contribution to the broadening effect, 
it is also possible to normalize the observable to heavier nuclei like helium, carbon, or 
neon, for which Fermi motion contribution is much larger than deuterium.

\subsection{Comparison to CLAS data}
A critical test for the model and its various components is to check whether it can describe data at different collision energies. To that purpose, we compare our 
model predictions, with unchanged parameters, to the multiplicity ratio for positively-charged 
pions measured by the CLAS Collaboration~\cite{CLAS:2021jhm}. The lower energy of the 
beam ($E_e=5.014$ GeV) implies a much lower range of lepton energy transfer variable 
$\nu$ as compared to the HERMES case, and 
the applicability of the TMD factorization and framework at such low energy on nuclei 
is questionable. Nevertheless, we found, in general, a satisfactory agreement between 
our model and the data, but some features are worthwhile to discuss.

\begin{figure*}
\begin{center}
\includegraphics[scale=0.6]{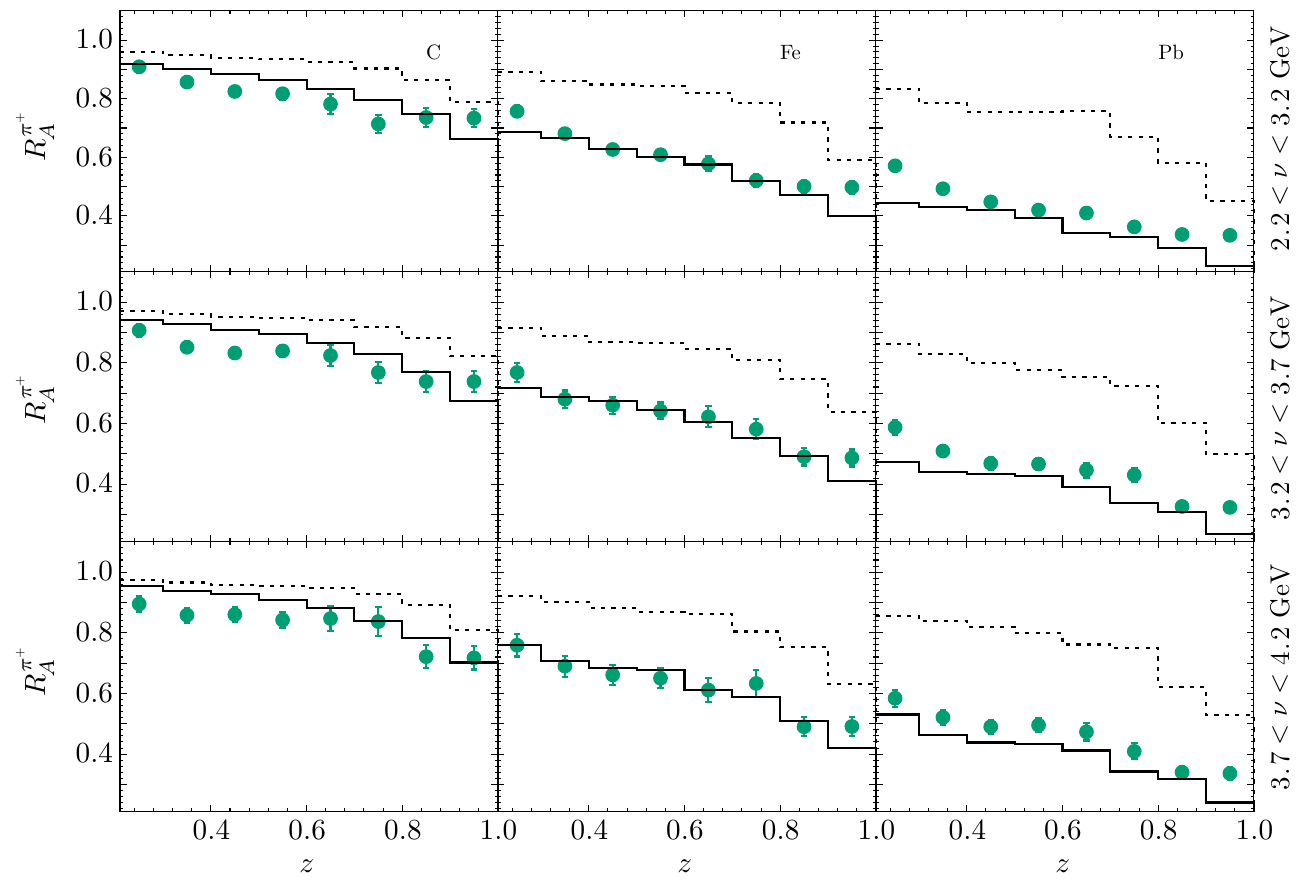}
\caption{$\pi^+$ multiplicity ratios as a function of $z$ for three different nuclear targets integrated in bins of $\nu$ and in $1.3 < \mbox{Q}^2 < 1.8$ Ge$\mbox{V}^2$. Model predictions are shown with default values (solid lines) and with absorption switched off (dotted), \textsl{i.e.} setting $n_{att}=1$ in Eq.~(\ref{eq:absorption}). Data from CLAS Collaboration~\cite{CLAS:2021jhm}.}
\label{plot:CLAS_ptbroad_vs_q2}
\end{center}
\end{figure*}

In Fig.~(\ref{plot:CLAS_ptbroad_vs_q2}) we present the multi-dimensional multiplicity 
ratios as a function of $z$ for three $\nu$ ranges and three target nuclei.
For clarity, we did not show the different $Q^2$ bins in this figure since the dependence 
on this variable is very weak, 
a feature also reproduced by our model.
The simulations reproduces very well the $z$-shapes across different values of $\nu$ and $A$ while it  slightly overshoots 
the multiplicity ratio for the light target (carbon) and slightly undershoots the one 
on the heavier target (lead).
The data does not show a dependence on $\nu$ in contrast with HERMES results, a feature 
reproduced by our model in this energy range. In the same figure, we present our calculation 
with absorption switched off, \textsl{i.e.} setting $n_{att}=1$ in Eq.~(\ref{eq:absorption}), 
in dotted lines. Comparing the two sets, one may observe that 
the impact of absorption for this data set 
is maximal and increasing in going from light to heavier nuclei, a direct consequence of 
the correlation introduced by the absorption mechanism in Eq.~(\ref{eq:absorption}).

\begin{figure*}
\begin{center}
\includegraphics[scale=0.55]{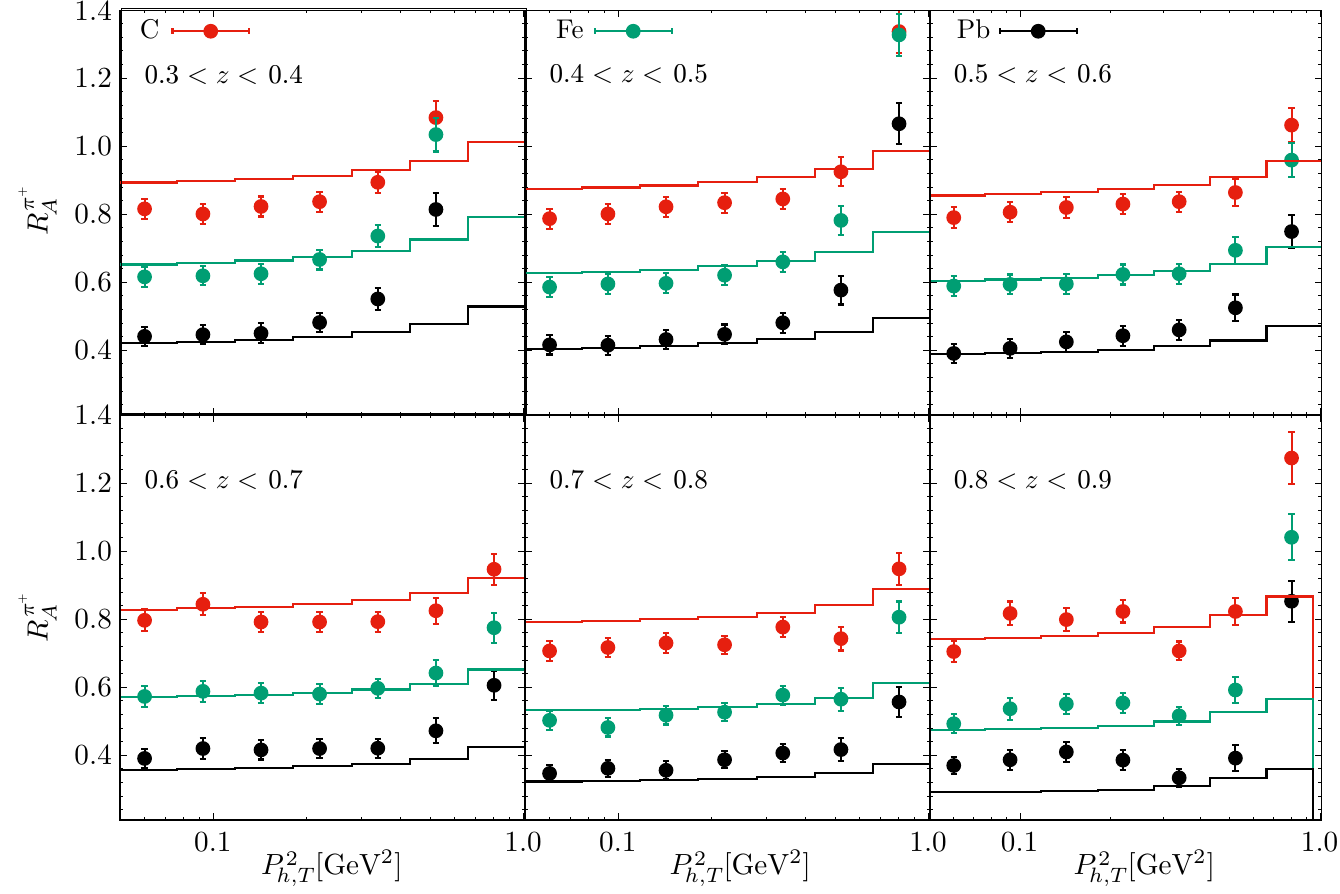}
\caption{Multiplicity ratios of $\pi^+$ as a function of $\pt^2$ in various $z$ bins for three different nuclear targets. Data from CLAS Collaboration~\cite{CLAS:2021jhm}.}
\label{plot:CLAS_ptbroad_vs_pt}
\end{center}
\end{figure*}

In Fig.~(\ref{plot:CLAS_ptbroad_vs_pt}), we present the multiplicity ratios as a function 
of the hadronic $\pt$ in slices of $z$ for different nuclear targets. Setting aside the 
visible normalization issue in some $z$ bins outlined above, we notice two things.
First, the slope of the multiplicity ratios as a function of $\pt$ is reduced as 
we go to higher $z$, which our model describes correctly at small $\pt$. 
Second, we are significantly undershooting the highest points in $\pt$. 
This could be a sign that the model generates an amount of broadening smaller than 
the one in the data, particularly with this discrepancy getting more pronounced going 
from light to heavy nuclei. 
However, it could also be the effect of reaching the end of 
the phase space in missing mass or due to the non-Gaussian nature at such high $\pt$.
In the absence of transverse momentum broadening results from CLAS, it is difficult to 
draw definite conclusions on this point. 

\subsection{Comparison to EMC data}
We finally compare our model to the multiplicity ratios data from the EMC collaboration on 
a copper target~\cite{EuropeanMuon:1991jmx}. The collected data refer to multiplicity 
ratios for the sum of charged hadrons, which we approximate as the sum of charged pions 
since the latter largely dominates the hadronic yields. This data set is particularly handy
since it allows us to test our model in the transition region from intermediate values of 
$\nu \approx 20$ GeV up to the asymptotic limit, where nuclear effects vanish.

\begin{figure*}
\begin{center}
\includegraphics[scale=0.6]{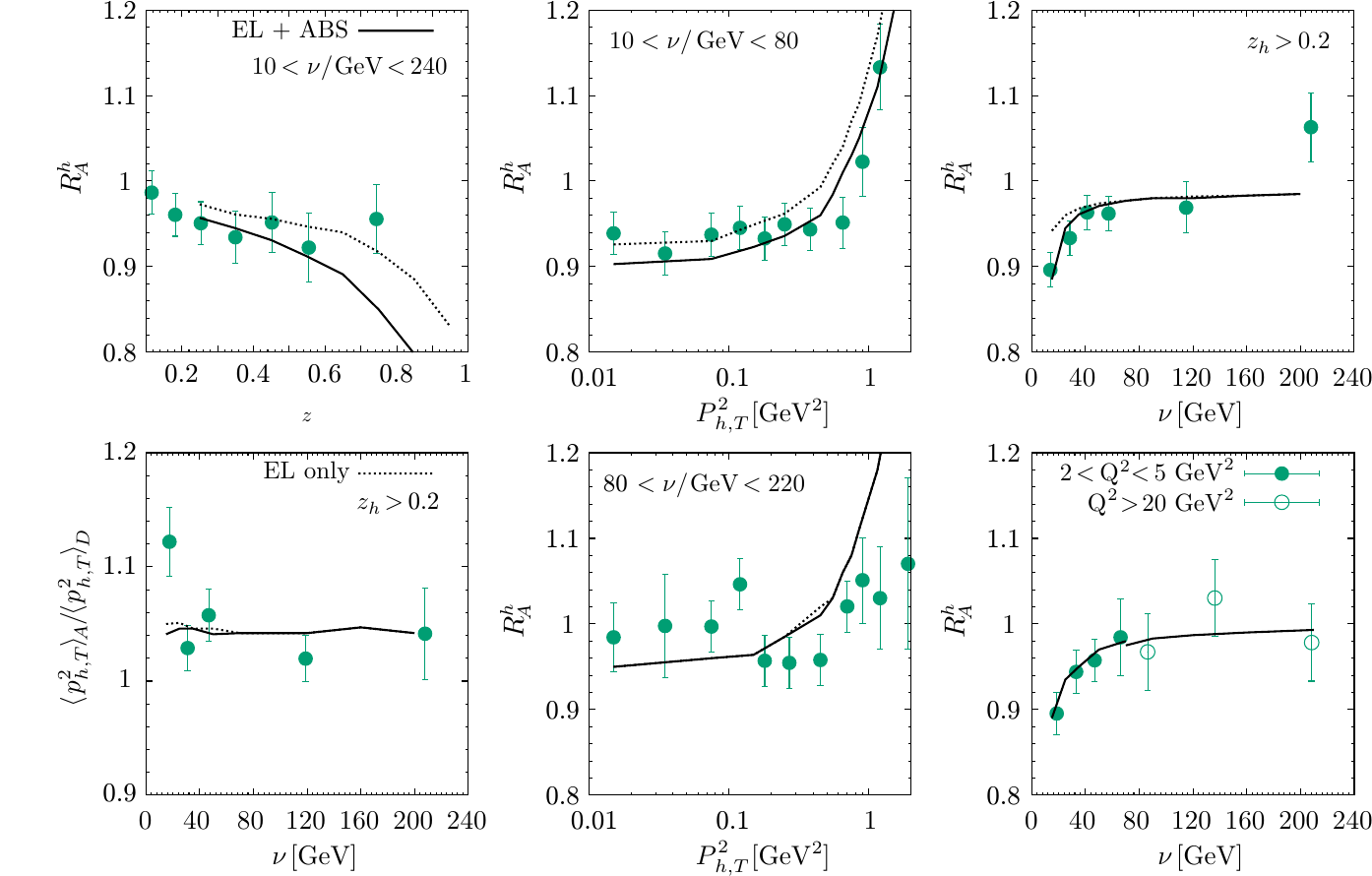}
\caption{Charged hadrons multiplicity ratios on copper as a function of kinematic variables compared to data from the European Muon Collaboration~\cite{EuropeanMuon:1991jmx}. 
Solid lines are for the full model with energy loss plus absorption, while dashed ones are for the model with absorption switched off by setting $n_{att}=1$. The bottom left panel shows the ratio of hadronic average transverse momentum on the nucleus over that on the deuterium.}
\label{plot:EMCratios}
\end{center}
\end{figure*}

In Fig.~(\ref{plot:EMCratios}), we present our model compared to data. The overall description 
is very good, particularly the shape of multiplicity ratios as a function of $z$ and $\nu$. 
We also present the calculation with absorption switched off (dotted lines): the 
comparison with the full model shows that, given the larger energy transfer, absorption has 
a much weaker impact on this data set. Nevertheless, absorption still gives a sizeable 
correction at the lower $\nu$ and is necessary to match the normalization and shape 
observed in the data. We conclude that the energy extrapolation is modeled correctly 
and reproduces the vanishing of nuclear effects at high energies.

\section{Conclusions}

We applied a TMD-based model to describe the nuclear SIDIS data of the CLAS, HERMES, and 
EMC collaborations, with the aim of a simultaneous description of multiplicity ratios and 
transverse momentum broadening. We found that using a model with parton energy loss alone 
leads to inconsistencies in the combined description of those observables, which can be 
achieved by adding a pre-hadron absorption mechanism instead. After tuning the model 
parameters, we found a satisfactory agreement with all the data sets considered.
We best describe the data with a cold nuclear matter transport coefficient value 
$\hat{q} =$ 0.3 GeV/fm$^2$, but this value is strongly correlated with the 
parametrization of both the absorption mechanism and the production length $L_P$. 

We find strong correlations between the observables and, in particular, an important 
effect of survivor bias on the transverse momentum broadening measurement. 
This demonstrates the importance of simultaneously describing both observables to
extract the transport coefficient, and more generally, to study the transverse 
effects in nuclei and understand hadronization off nuclei in SIDIS.
The model describes the data on a very wide range of energies from $\nu =2$~GeV to 
$\nu =200$~GeV. We find that at low lepton energy transfer, hadron absorption is 
responsible for the majority of the depletion observed in multiplicity ratios, while 
at higher lepton energy transfer, parton energy loss becomes relatively more important, 
until the smooth vanishing of all nuclear effects at the highest energies.
The model finally presents problems describing the correct $A$ and $z$ dependencies 
of the transverse momentum broadening. This
issue definitely requires further investigations on both the experimental and 
theoretical front to be resolved.

\section{Acknowledgements}
We acknowledge many fruitful discussions with A.~Accardi, F.~Arleo, and S.~Peigne.
This work has received funding from the European Research Council (ERC) under the
European Union’s Horizon 2020 research and innovation programme (Grant agreement No.
804480)

\vspace{0.5cm}

\appendix
\section{Hadronic widths} 
\label{appx}
\begin{figure*}
\begin{center}
\includegraphics[scale=0.5]{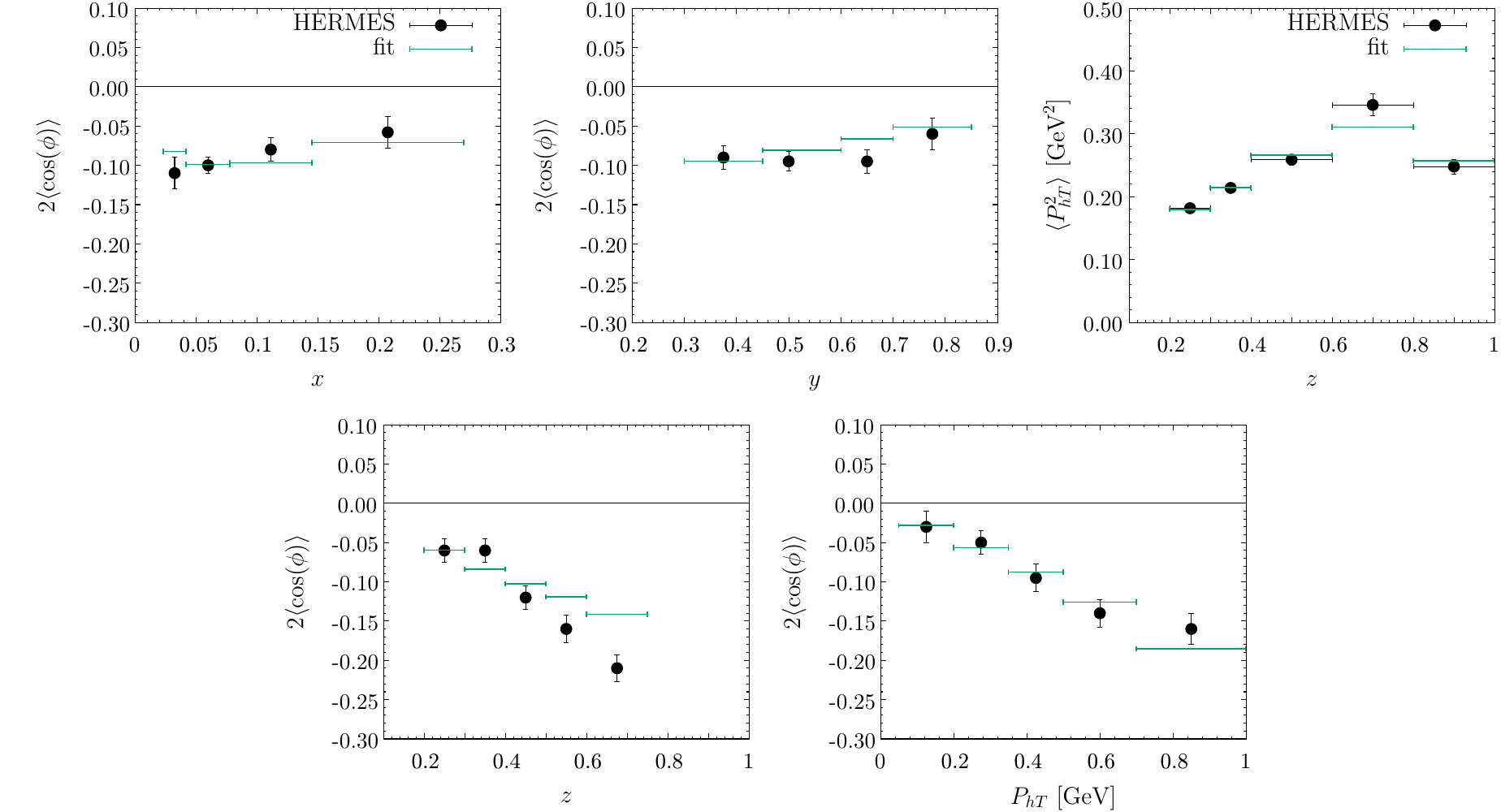}
\caption{Predictions based on best fit results compared to the $\cos(\phi)$ asymmetry \cite{HERMES:2012kpt} and the hadronic widths  \cite{HERMES:2012uyd} measured in $\pi^+$ electroproduction on a proton target at HERMES.}
\label{fit}
\end{center}
\end{figure*}

\noindent
Numerous extraction of parameters characterizing transverse momentum spectra within TMD phenomenology are available in the literature. In the present work, however,  
we wish to provide a dedicated extraction within the specific theoretical framework adopted and tailored to the analysis of the nuclear dataset. 
For that purpose, we present in the right panel of Fig.~(\ref{fit}) the hadronic width as a function of $z$. On the other hand, the 
width parametrization in Eq.~(\ref{eq:meanpt}) shows a quadratic behaviour as a function of $z$, at variance with the data which 
shows a dampening at large $z$. To accommodate this behavior within the structure of Eq.~(\ref{eq:meanpt}) and considering the findings of Ref.[28,29], 
we assume a fragmenting width of the type 
$\langle p_\perp^2 \rangle=(A+B z^2)(1-z)^C$ and 
a constant, $z$-independent, distribution width 
$\langle k_\perp^2 \rangle=D$. 
The four-parameter functional form is then given by
\begin{equation}
\langle \pt^2 \rangle=(A+B z^2)(1-z)^C+z^2 D\,.
\end{equation}
In such a form, however, the $D$ parameter can not be precisely determined by fits to hadronic spectra alone. We therefore reconsider the cross section in Eq.~(\ref{e:crossmaster}). There, the structure function proportional to $\cos \phi$, the so-called Cahn contribution, can be written, under the approximation used in this paper, as 
\begin{equation}
F_{UU}^{\cos \phi_h}=-2 \xbj \frac{\pt}{Q }\sum_q  e_q^2 f_{q/P}(\xbj,Q^2) \> D_{h/q}(z,Q^2) \frac{z \langle k_\perp^2\rangle}{\langle\pt^2\rangle}
\frac{e^{-\pt^2/\langle\pt^2\rangle}}{\pi\langle\pt^2\rangle}\,.
\label{FUUcosphi}
\end{equation}
in whose terms one can then define the following asymmetry 
\begin{equation}
2 \langle \cos \phi \rangle = 2\frac{\int d\phi d\sigma \cos \phi}{\int d\phi d\sigma}=\frac{2(2-y)\sqrt{1-y} F_{UU}^{\cos \phi}}{[1+(1-y)^2] F_{UU}}\,. 
\end{equation}
The latter does not contain additional unknowns with respect to 
$F_{UU}$ and, important for our purpose, is proportional to $\langle k_\perp^2 \rangle$: its inclusion in the fit allows us to constrain the parameter $D$, which would be otherwise under-unconstrained. 
The parameters are then obtained by a minimization procedure performed with the help of the \texttt{MINUIT} package \cite{James:1975dr}. We include in the fit the widths extracted by Gaussian fits of $\pi^+$ transverse momentum spectra off a proton target and the corresponding $\cos(\phi)$ asymmetry, both published by the HERMES Collaboration \cite{HERMES:2012uyd,HERMES:2012kpt}. 
We assume that the asymmetry is saturated by the Cahn effect, and the Boer–Mulders contribution is negligible. The best-fit results are compared to data in Fig.~(\ref{fit}) and the corresponding parameters are reported in Tab.~(\ref{tab:widths}).
The fit returns a $\chi^2$ per \textsl{d.o.f.} of around 2, and its large value is traced back to the discrepancy observed in the description of the asymmetry as a function of $z$. Despite this, the obtained parametrization gives a good representation of $ \langle \pt^2 \rangle$, as shown in the right panel of Fig.~(\ref{fit}), and therefore is used as a default choice in all our simulations.

\begin{table}
\setlength\extrarowheight{2pt}
\begin{center}
\begin{tabular}{ P{4cm} P{4cm} P{3cm} P{4cm} } 
\hline \hline 
   A       &  B &  C & D  \\ 
\hline 
   0.15  $\pm$   0.01 $\mbox{GeV}^2$   &     1.08  $\pm$   0.15 $\mbox{GeV}^2$   &    0.69  $\pm$   0.07  &    0.052 $\pm$   0.002  $\mbox{GeV}^2$   \\ 
\hline   
\end{tabular}
\end{center}
\caption{Parameters controlling the hadronic widths on a proton target obtained from the fit of HERMES data.} 
\label{tab:widths}
\end{table}

\bibliographystyle{apsrev4-1}
\bibliography{nTMD_biblio.bib}

\begin{thebibliography}{41}%
\makeatletter
\providecommand \@ifxundefined [1]{%
 \@ifx{#1\undefined}
}%
\providecommand \@ifnum [1]{%
 \ifnum #1\expandafter \@firstoftwo
 \else \expandafter \@secondoftwo
 \fi
}%
\providecommand \@ifx [1]{%
 \ifx #1\expandafter \@firstoftwo
 \else \expandafter \@secondoftwo
 \fi
}%
\providecommand \natexlab [1]{#1}%
\providecommand \enquote  [1]{``#1''}%
\providecommand \bibnamefont  [1]{#1}%
\providecommand \bibfnamefont [1]{#1}%
\providecommand \citenamefont [1]{#1}%
\providecommand \href@noop [0]{\@secondoftwo}%
\providecommand \href [0]{\begingroup \@sanitize@url \@href}%
\providecommand \@href[1]{\@@startlink{#1}\@@href}%
\providecommand \@@href[1]{\endgroup#1\@@endlink}%
\providecommand \@sanitize@url [0]{\catcode `\\12\catcode `\$12\catcode
  `\&12\catcode `\#12\catcode `\^12\catcode `\_12\catcode `\%12\relax}%
\providecommand \@@startlink[1]{}%
\providecommand \@@endlink[0]{}%
\providecommand \url  [0]{\begingroup\@sanitize@url \@url }%
\providecommand \@url [1]{\endgroup\@href {#1}{\urlprefix }}%
\providecommand \urlprefix  [0]{URL }%
\providecommand \Eprint [0]{\href }%
\providecommand \doibase [0]{http://dx.doi.org/}%
\providecommand \selectlanguage [0]{\@gobble}%
\providecommand \bibinfo  [0]{\@secondoftwo}%
\providecommand \bibfield  [0]{\@secondoftwo}%
\providecommand \translation [1]{[#1]}%
\providecommand \BibitemOpen [0]{}%
\providecommand \bibitemStop [0]{}%
\providecommand \bibitemNoStop [0]{.\EOS\space}%
\providecommand \EOS [0]{\spacefactor3000\relax}%
\providecommand \BibitemShut  [1]{\csname bibitem#1\endcsname}%
\let\auto@bib@innerbib\@empty
\bibitem [{\citenamefont {Workman}\ \emph {et~al.}(2022)\citenamefont {Workman}
  \emph {et~al.}}]{ParticleDataGroup:2022pthFF}%
  \BibitemOpen
  \bibfield  {author} {\bibinfo {author} {\bibfnamefont {R.~L.}\ \bibnamefont
  {Workman}} \emph {et~al.} (\bibinfo {collaboration} {Particle Data Group}),\
  }\enquote {\bibinfo {title} {{Fragmentation Functions in $e^+e^-$, $ep$, and
  $pp$ Collisions}},}\ in\ \href {\doibase 10.1093/ptep/ptac097} {\emph
  {\bibinfo {booktitle} {{Prog. Theor. Exp. Phys. 2022, 083C01}}}}\ (\bibinfo
  {year} {2022})\ Chap.~\bibinfo {chapter} {19}, p.\ \bibinfo {pages}
  {375}\BibitemShut {NoStop}%
\bibitem [{\citenamefont {Accardi}\ \emph {et~al.}(2010)\citenamefont
  {Accardi}, \citenamefont {Arleo}, \citenamefont {Brooks}, \citenamefont
  {D'Enterria},\ and\ \citenamefont {Muccifora}}]{Accardi:2009qv}%
  \BibitemOpen
  \bibfield  {author} {\bibinfo {author} {\bibfnamefont {A.}~\bibnamefont
  {Accardi}}, \bibinfo {author} {\bibfnamefont {F.}~\bibnamefont {Arleo}},
  \bibinfo {author} {\bibfnamefont {W.~K.}\ \bibnamefont {Brooks}}, \bibinfo
  {author} {\bibfnamefont {D.}~\bibnamefont {D'Enterria}}, \ and\ \bibinfo
  {author} {\bibfnamefont {V.}~\bibnamefont {Muccifora}},\ }\href {\doibase
  10.1393/ncr/i2009-10048-0} {\bibfield  {journal} {\bibinfo  {journal} {Riv.
  Nuovo Cim.}\ }\textbf {\bibinfo {volume} {32}},\ \bibinfo {pages} {439}
  (\bibinfo {year} {2010})},\ \Eprint {http://arxiv.org/abs/0907.3534}
  {arXiv:0907.3534 [nucl-th]} \BibitemShut {NoStop}%
\bibitem [{\citenamefont {Avakian}\ \emph {et~al.}(2016)\citenamefont
  {Avakian}, \citenamefont {Bressan},\ and\ \citenamefont
  {Contalbrigo}}]{Avakian:2016rst}%
  \BibitemOpen
  \bibfield  {author} {\bibinfo {author} {\bibfnamefont {H.}~\bibnamefont
  {Avakian}}, \bibinfo {author} {\bibfnamefont {A.}~\bibnamefont {Bressan}}, \
  and\ \bibinfo {author} {\bibfnamefont {M.}~\bibnamefont {Contalbrigo}},\
  }\href {\doibase 10.1140/epja/i2016-16150-x} {\bibfield  {journal} {\bibinfo
  {journal} {Eur. Phys. J. A}\ }\textbf {\bibinfo {volume} {52}},\ \bibinfo
  {pages} {150} (\bibinfo {year} {2016})},\ \bibinfo {note} {[Erratum:
  Eur.Phys.J.A 52, 165 (2016)]}\BibitemShut {NoStop}%
\bibitem [{\citenamefont {Bacchetta}(2016)}]{Bacchetta:2016ccz}%
  \BibitemOpen
  \bibfield  {author} {\bibinfo {author} {\bibfnamefont {A.}~\bibnamefont
  {Bacchetta}},\ }\href {\doibase 10.1140/epja/i2016-16163-5} {\bibfield
  {journal} {\bibinfo  {journal} {Eur. Phys. J. A}\ }\textbf {\bibinfo {volume}
  {52}},\ \bibinfo {pages} {163} (\bibinfo {year} {2016})},\ \Eprint
  {http://arxiv.org/abs/2107.06772} {arXiv:2107.06772 [hep-ph]} \BibitemShut
  {NoStop}%
\bibitem [{\citenamefont {Alrashed}\ \emph {et~al.}(2022)\citenamefont
  {Alrashed}, \citenamefont {Anderle}, \citenamefont {Kang}, \citenamefont
  {Terry},\ and\ \citenamefont {Xing}}]{Alrashed:2021csd}%
  \BibitemOpen
  \bibfield  {author} {\bibinfo {author} {\bibfnamefont {M.}~\bibnamefont
  {Alrashed}}, \bibinfo {author} {\bibfnamefont {D.}~\bibnamefont {Anderle}},
  \bibinfo {author} {\bibfnamefont {Z.-B.}\ \bibnamefont {Kang}}, \bibinfo
  {author} {\bibfnamefont {J.}~\bibnamefont {Terry}}, \ and\ \bibinfo {author}
  {\bibfnamefont {H.}~\bibnamefont {Xing}},\ }\href {\doibase
  10.1103/PhysRevLett.129.242001} {\bibfield  {journal} {\bibinfo  {journal}
  {Phys. Rev. Lett.}\ }\textbf {\bibinfo {volume} {129}},\ \bibinfo {pages}
  {242001} (\bibinfo {year} {2022})},\ \Eprint
  {http://arxiv.org/abs/2107.12401} {arXiv:2107.12401 [hep-ph]} \BibitemShut
  {NoStop}%
\bibitem [{\citenamefont {Ashman}\ \emph {et~al.}(1991)\citenamefont {Ashman}
  \emph {et~al.}}]{EuropeanMuon:1991jmx}%
  \BibitemOpen
  \bibfield  {author} {\bibinfo {author} {\bibfnamefont {J.}~\bibnamefont
  {Ashman}} \emph {et~al.} (\bibinfo {collaboration} {European Muon
  Collaboration}),\ }\href {\doibase 10.1007/BF01412322} {\bibfield  {journal}
  {\bibinfo  {journal} {Z. Phys. C}\ }\textbf {\bibinfo {volume} {52}},\
  \bibinfo {pages} {1} (\bibinfo {year} {1991})}\BibitemShut {NoStop}%
\bibitem [{\citenamefont {Airapetian}\ \emph {et~al.}(2007)\citenamefont
  {Airapetian} \emph {et~al.}}]{Airapetian:2007vu}%
  \BibitemOpen
  \bibfield  {author} {\bibinfo {author} {\bibfnamefont {A.}~\bibnamefont
  {Airapetian}} \emph {et~al.} (\bibinfo {collaboration} {HERMES
  Collaboration}),\ }\href {\doibase 10.1016/j.nuclphysb.2007.06.004}
  {\bibfield  {journal} {\bibinfo  {journal} {Nucl.Phys.}\ }\textbf {\bibinfo
  {volume} {B780}},\ \bibinfo {pages} {1} (\bibinfo {year} {2007})},\ \Eprint
  {http://arxiv.org/abs/0704.3270} {arXiv:0704.3270 [hep-ex]} \BibitemShut
  {NoStop}%
\bibitem [{\citenamefont {Airapetian}\ \emph {et~al.}(2010)\citenamefont
  {Airapetian} \emph {et~al.}}]{HERMES:2009uge}%
  \BibitemOpen
  \bibfield  {author} {\bibinfo {author} {\bibfnamefont {A.}~\bibnamefont
  {Airapetian}} \emph {et~al.} (\bibinfo {collaboration} {HERMES}),\ }\href
  {\doibase 10.1016/j.physletb.2010.01.020} {\bibfield  {journal} {\bibinfo
  {journal} {Phys. Lett. B}\ }\textbf {\bibinfo {volume} {684}},\ \bibinfo
  {pages} {114} (\bibinfo {year} {2010})},\ \Eprint
  {http://arxiv.org/abs/0906.2478} {arXiv:0906.2478 [hep-ex]} \BibitemShut
  {NoStop}%
\bibitem [{\citenamefont {Airapetian}\ \emph {et~al.}(2011)\citenamefont
  {Airapetian} \emph {et~al.}}]{HERMES:2011qjb}%
  \BibitemOpen
  \bibfield  {author} {\bibinfo {author} {\bibfnamefont {A.}~\bibnamefont
  {Airapetian}} \emph {et~al.} (\bibinfo {collaboration} {HERMES}),\ }\href
  {\doibase 10.1140/epja/i2011-11113-5} {\bibfield  {journal} {\bibinfo
  {journal} {Eur. Phys. J. A}\ }\textbf {\bibinfo {volume} {47}},\ \bibinfo
  {pages} {113} (\bibinfo {year} {2011})},\ \Eprint
  {http://arxiv.org/abs/1107.3496} {arXiv:1107.3496 [hep-ex]} \BibitemShut
  {NoStop}%
\bibitem [{\citenamefont {Moran}\ \emph {et~al.}(2022)\citenamefont {Moran}
  \emph {et~al.}}]{CLAS:2021jhm}%
  \BibitemOpen
  \bibfield  {author} {\bibinfo {author} {\bibfnamefont {S.}~\bibnamefont
  {Moran}} \emph {et~al.} (\bibinfo {collaboration} {CLAS}),\ }\href {\doibase
  10.1103/PhysRevC.105.015201} {\bibfield  {journal} {\bibinfo  {journal}
  {Phys. Rev. C}\ }\textbf {\bibinfo {volume} {105}},\ \bibinfo {pages}
  {015201} (\bibinfo {year} {2022})},\ \Eprint
  {http://arxiv.org/abs/2109.09951} {arXiv:2109.09951 [nucl-ex]} \BibitemShut
  {NoStop}%
\bibitem [{\citenamefont {Kopeliovich}\ \emph {et~al.}(2004)\citenamefont
  {Kopeliovich}, \citenamefont {Nemchik}, \citenamefont {Predazzi},\ and\
  \citenamefont {Hayashigaki}}]{Kopeliovich:2003py}%
  \BibitemOpen
  \bibfield  {author} {\bibinfo {author} {\bibfnamefont {B.}~\bibnamefont
  {Kopeliovich}}, \bibinfo {author} {\bibfnamefont {J.}~\bibnamefont
  {Nemchik}}, \bibinfo {author} {\bibfnamefont {E.}~\bibnamefont {Predazzi}}, \
  and\ \bibinfo {author} {\bibfnamefont {A.}~\bibnamefont {Hayashigaki}},\
  }\href {\doibase 10.1016/j.nuclphysa.2004.04.110} {\bibfield  {journal}
  {\bibinfo  {journal} {Nucl.Phys.}\ }\textbf {\bibinfo {volume} {A740}},\
  \bibinfo {pages} {211} (\bibinfo {year} {2004})},\ \Eprint
  {http://arxiv.org/abs/hep-ph/0311220} {arXiv:hep-ph/0311220 [hep-ph]}
  \BibitemShut {NoStop}%
\bibitem [{\citenamefont {Accardi}\ \emph {et~al.}(2003)\citenamefont
  {Accardi}, \citenamefont {Muccifora},\ and\ \citenamefont
  {Pirner}}]{Accardi:2002tv}%
  \BibitemOpen
  \bibfield  {author} {\bibinfo {author} {\bibfnamefont {A.}~\bibnamefont
  {Accardi}}, \bibinfo {author} {\bibfnamefont {V.}~\bibnamefont {Muccifora}},
  \ and\ \bibinfo {author} {\bibfnamefont {H.-J.}\ \bibnamefont {Pirner}},\
  }\href {\doibase 10.1016/S0375-9474(03)00670-5} {\bibfield  {journal}
  {\bibinfo  {journal} {Nucl.Phys.}\ }\textbf {\bibinfo {volume} {A720}},\
  \bibinfo {pages} {131} (\bibinfo {year} {2003})},\ \Eprint
  {http://arxiv.org/abs/nucl-th/0211011} {arXiv:nucl-th/0211011 [nucl-th]}
  \BibitemShut {NoStop}%
\bibitem [{\citenamefont {Accardi}\ \emph {et~al.}(2005)\citenamefont
  {Accardi}, \citenamefont {Grunewald}, \citenamefont {Muccifora},\ and\
  \citenamefont {Pirner}}]{Accardi:2005hk}%
  \BibitemOpen
  \bibfield  {author} {\bibinfo {author} {\bibfnamefont {A.}~\bibnamefont
  {Accardi}}, \bibinfo {author} {\bibfnamefont {D.}~\bibnamefont {Grunewald}},
  \bibinfo {author} {\bibfnamefont {V.}~\bibnamefont {Muccifora}}, \ and\
  \bibinfo {author} {\bibfnamefont {H.~J.}\ \bibnamefont {Pirner}},\ }\href
  {\doibase 10.1016/j.nuclphysa.2005.05.201} {\bibfield  {journal} {\bibinfo
  {journal} {Nucl. Phys. A}\ }\textbf {\bibinfo {volume} {761}},\ \bibinfo
  {pages} {67} (\bibinfo {year} {2005})},\ \Eprint
  {http://arxiv.org/abs/hep-ph/0502072} {arXiv:hep-ph/0502072} \BibitemShut
  {NoStop}%
\bibitem [{\citenamefont {Guiot}\ and\ \citenamefont
  {Kopeliovich}(2020)}]{Guiot:2020vsf}%
  \BibitemOpen
  \bibfield  {author} {\bibinfo {author} {\bibfnamefont {B.}~\bibnamefont
  {Guiot}}\ and\ \bibinfo {author} {\bibfnamefont {B.~Z.}\ \bibnamefont
  {Kopeliovich}},\ }\href {\doibase 10.1103/PhysRevC.102.045201} {\bibfield
  {journal} {\bibinfo  {journal} {Phys. Rev. C}\ }\textbf {\bibinfo {volume}
  {102}},\ \bibinfo {pages} {045201} (\bibinfo {year} {2020})},\ \Eprint
  {http://arxiv.org/abs/2001.00974} {arXiv:2001.00974 [hep-ph]} \BibitemShut
  {NoStop}%
\bibitem [{\citenamefont {Arleo}(2003)}]{Arleo:2003jz}%
  \BibitemOpen
  \bibfield  {author} {\bibinfo {author} {\bibfnamefont {F.}~\bibnamefont
  {Arleo}},\ }\href {\doibase 10.1140/epjc/s2003-01289-x} {\bibfield  {journal}
  {\bibinfo  {journal} {Eur. Phys. J. C}\ }\textbf {\bibinfo {volume} {30}},\
  \bibinfo {pages} {213} (\bibinfo {year} {2003})},\ \Eprint
  {http://arxiv.org/abs/hep-ph/0306235} {arXiv:hep-ph/0306235} \BibitemShut
  {NoStop}%
\bibitem [{\citenamefont {Domdey}\ \emph {et~al.}(2009)\citenamefont {Domdey},
  \citenamefont {Grunewald}, \citenamefont {Kopeliovich},\ and\ \citenamefont
  {Pirner}}]{Domdey:2008aq}%
  \BibitemOpen
  \bibfield  {author} {\bibinfo {author} {\bibfnamefont {S.}~\bibnamefont
  {Domdey}}, \bibinfo {author} {\bibfnamefont {D.}~\bibnamefont {Grunewald}},
  \bibinfo {author} {\bibfnamefont {B.}~\bibnamefont {Kopeliovich}}, \ and\
  \bibinfo {author} {\bibfnamefont {H.}~\bibnamefont {Pirner}},\ }\href
  {\doibase 10.1016/j.nuclphysa.2009.04.009} {\bibfield  {journal} {\bibinfo
  {journal} {Nucl.Phys.}\ }\textbf {\bibinfo {volume} {A825}},\ \bibinfo
  {pages} {200} (\bibinfo {year} {2009})},\ \Eprint
  {http://arxiv.org/abs/0812.2838} {arXiv:0812.2838 [hep-ph]} \BibitemShut
  {NoStop}%
\bibitem [{\citenamefont {Gao}\ \emph {et~al.}(2010)\citenamefont {Gao},
  \citenamefont {Liang},\ and\ \citenamefont {Wang}}]{Gao:2010mj}%
  \BibitemOpen
  \bibfield  {author} {\bibinfo {author} {\bibfnamefont {J.-H.}\ \bibnamefont
  {Gao}}, \bibinfo {author} {\bibfnamefont {Z.-T.}\ \bibnamefont {Liang}}, \
  and\ \bibinfo {author} {\bibfnamefont {X.-N.}\ \bibnamefont {Wang}},\ }\href
  {\doibase 10.1103/PhysRevC.81.065211} {\bibfield  {journal} {\bibinfo
  {journal} {Phys. Rev. C}\ }\textbf {\bibinfo {volume} {81}},\ \bibinfo
  {pages} {065211} (\bibinfo {year} {2010})},\ \Eprint
  {http://arxiv.org/abs/1001.3146} {arXiv:1001.3146 [hep-ph]} \BibitemShut
  {NoStop}%
\bibitem [{\citenamefont {Song}\ and\ \citenamefont
  {Duan}(2010)}]{Song:2010zza}%
  \BibitemOpen
  \bibfield  {author} {\bibinfo {author} {\bibfnamefont {L.-H.}\ \bibnamefont
  {Song}}\ and\ \bibinfo {author} {\bibfnamefont {C.-G.}\ \bibnamefont
  {Duan}},\ }\href {\doibase 10.1103/PhysRevC.81.035207} {\bibfield  {journal}
  {\bibinfo  {journal} {Phys.Rev.}\ }\textbf {\bibinfo {volume} {C81}},\
  \bibinfo {pages} {035207} (\bibinfo {year} {2010})}\BibitemShut {NoStop}%
\bibitem [{\citenamefont {Song}\ \emph
  {et~al.}(2014{\natexlab{a}})\citenamefont {Song}, \citenamefont {Gao},
  \citenamefont {Liang},\ and\ \citenamefont {Wang}}]{Song:2013sja}%
  \BibitemOpen
  \bibfield  {author} {\bibinfo {author} {\bibfnamefont {Y.-K.}\ \bibnamefont
  {Song}}, \bibinfo {author} {\bibfnamefont {J.-H.}\ \bibnamefont {Gao}},
  \bibinfo {author} {\bibfnamefont {Z.-T.}\ \bibnamefont {Liang}}, \ and\
  \bibinfo {author} {\bibfnamefont {X.-N.}\ \bibnamefont {Wang}},\ }\href
  {\doibase 10.1103/PhysRevD.89.014005} {\bibfield  {journal} {\bibinfo
  {journal} {Phys. Rev. D}\ }\textbf {\bibinfo {volume} {89}},\ \bibinfo
  {pages} {014005} (\bibinfo {year} {2014}{\natexlab{a}})},\ \Eprint
  {http://arxiv.org/abs/1308.1159} {arXiv:1308.1159 [hep-ph]} \BibitemShut
  {NoStop}%
\bibitem [{\citenamefont {Song}\ \emph
  {et~al.}(2014{\natexlab{b}})\citenamefont {Song}, \citenamefont {Liang},\
  and\ \citenamefont {Wang}}]{Song:2014sja}%
  \BibitemOpen
  \bibfield  {author} {\bibinfo {author} {\bibfnamefont {Y.-K.}\ \bibnamefont
  {Song}}, \bibinfo {author} {\bibfnamefont {Z.-T.}\ \bibnamefont {Liang}}, \
  and\ \bibinfo {author} {\bibfnamefont {X.-N.}\ \bibnamefont {Wang}},\ }\href
  {\doibase 10.1103/PhysRevD.89.117501} {\bibfield  {journal} {\bibinfo
  {journal} {Phys. Rev. D}\ }\textbf {\bibinfo {volume} {89}},\ \bibinfo
  {pages} {117501} (\bibinfo {year} {2014}{\natexlab{b}})},\ \Eprint
  {http://arxiv.org/abs/1402.3042} {arXiv:1402.3042 [nucl-th]} \BibitemShut
  {NoStop}%
\bibitem [{\citenamefont {Ru}\ \emph {et~al.}(2021)\citenamefont {Ru},
  \citenamefont {Kang}, \citenamefont {Wang}, \citenamefont {Xing},\ and\
  \citenamefont {Zhang}}]{Ru:2019qvz}%
  \BibitemOpen
  \bibfield  {author} {\bibinfo {author} {\bibfnamefont {P.}~\bibnamefont
  {Ru}}, \bibinfo {author} {\bibfnamefont {Z.-B.}\ \bibnamefont {Kang}},
  \bibinfo {author} {\bibfnamefont {E.}~\bibnamefont {Wang}}, \bibinfo {author}
  {\bibfnamefont {H.}~\bibnamefont {Xing}}, \ and\ \bibinfo {author}
  {\bibfnamefont {B.-W.}\ \bibnamefont {Zhang}},\ }\href {\doibase
  10.1103/PhysRevD.103.L031901} {\bibfield  {journal} {\bibinfo  {journal}
  {Phys. Rev. D}\ }\textbf {\bibinfo {volume} {103}},\ \bibinfo {pages}
  {L031901} (\bibinfo {year} {2021})},\ \Eprint
  {http://arxiv.org/abs/1907.11808} {arXiv:1907.11808 [hep-ph]} \BibitemShut
  {NoStop}%
\bibitem [{\citenamefont {Brooks}\ and\ \citenamefont
  {L\'opez}(2021)}]{Brooks:2020fmf}%
  \BibitemOpen
  \bibfield  {author} {\bibinfo {author} {\bibfnamefont {W.~K.}\ \bibnamefont
  {Brooks}}\ and\ \bibinfo {author} {\bibfnamefont {J.~A.}\ \bibnamefont
  {L\'opez}},\ }\href {\doibase 10.1016/j.physletb.2021.136171} {\bibfield
  {journal} {\bibinfo  {journal} {Phys. Lett. B}\ }\textbf {\bibinfo {volume}
  {816}},\ \bibinfo {pages} {136171} (\bibinfo {year} {2021})},\ \Eprint
  {http://arxiv.org/abs/2004.07236} {arXiv:2004.07236 [hep-ph]} \BibitemShut
  {NoStop}%
\bibitem [{\citenamefont {Gallmeister}\ and\ \citenamefont
  {Mosel}(2008)}]{Gallmeister:2007an}%
  \BibitemOpen
  \bibfield  {author} {\bibinfo {author} {\bibfnamefont {K.}~\bibnamefont
  {Gallmeister}}\ and\ \bibinfo {author} {\bibfnamefont {U.}~\bibnamefont
  {Mosel}},\ }\href {\doibase 10.1016/j.nuclphysa.2007.12.009} {\bibfield
  {journal} {\bibinfo  {journal} {Nucl. Phys. A}\ }\textbf {\bibinfo {volume}
  {801}},\ \bibinfo {pages} {68} (\bibinfo {year} {2008})},\ \Eprint
  {http://arxiv.org/abs/nucl-th/0701064} {arXiv:nucl-th/0701064} \BibitemShut
  {NoStop}%
\bibitem [{\citenamefont {Ke}\ \emph {et~al.}(2023)\citenamefont {Ke},
  \citenamefont {Zhang}, \citenamefont {Xing},\ and\ \citenamefont
  {Wang}}]{Ke:2023xeo}%
  \BibitemOpen
  \bibfield  {author} {\bibinfo {author} {\bibfnamefont {W.}~\bibnamefont
  {Ke}}, \bibinfo {author} {\bibfnamefont {Y.-Y.}\ \bibnamefont {Zhang}},
  \bibinfo {author} {\bibfnamefont {H.}~\bibnamefont {Xing}}, \ and\ \bibinfo
  {author} {\bibfnamefont {X.-N.}\ \bibnamefont {Wang}},\ }\href@noop {} {\
  (\bibinfo {year} {2023})},\ \Eprint {http://arxiv.org/abs/2304.10779}
  {arXiv:2304.10779 [hep-ph]} \BibitemShut {NoStop}%
\bibitem [{\citenamefont {Bacchetta}\ \emph {et~al.}(2004)\citenamefont
  {Bacchetta}, \citenamefont {D'Alesio}, \citenamefont {Diehl},\ and\
  \citenamefont {Miller}}]{Bacchetta:2004jz}%
  \BibitemOpen
  \bibfield  {author} {\bibinfo {author} {\bibfnamefont {A.}~\bibnamefont
  {Bacchetta}}, \bibinfo {author} {\bibfnamefont {U.}~\bibnamefont {D'Alesio}},
  \bibinfo {author} {\bibfnamefont {M.}~\bibnamefont {Diehl}}, \ and\ \bibinfo
  {author} {\bibfnamefont {C.~A.}\ \bibnamefont {Miller}},\ }\href {\doibase
  10.1103/PhysRevD.70.117504} {\bibfield  {journal} {\bibinfo  {journal} {Phys.
  Rev. D}\ }\textbf {\bibinfo {volume} {70}},\ \bibinfo {pages} {117504}
  (\bibinfo {year} {2004})},\ \Eprint {http://arxiv.org/abs/hep-ph/0410050}
  {arXiv:hep-ph/0410050} \BibitemShut {NoStop}%
\bibitem [{\citenamefont {Bacchetta}\ \emph {et~al.}(2007)\citenamefont
  {Bacchetta}, \citenamefont {Diehl}, \citenamefont {Goeke}, \citenamefont
  {Metz}, \citenamefont {Mulders},\ and\ \citenamefont
  {Schlegel}}]{Bacchetta:2006tn}%
  \BibitemOpen
  \bibfield  {author} {\bibinfo {author} {\bibfnamefont {A.}~\bibnamefont
  {Bacchetta}}, \bibinfo {author} {\bibfnamefont {M.}~\bibnamefont {Diehl}},
  \bibinfo {author} {\bibfnamefont {K.}~\bibnamefont {Goeke}}, \bibinfo
  {author} {\bibfnamefont {A.}~\bibnamefont {Metz}}, \bibinfo {author}
  {\bibfnamefont {P.~J.}\ \bibnamefont {Mulders}}, \ and\ \bibinfo {author}
  {\bibfnamefont {M.}~\bibnamefont {Schlegel}},\ }\href {\doibase
  10.1088/1126-6708/2007/02/093} {\bibfield  {journal} {\bibinfo  {journal}
  {JHEP}\ }\textbf {\bibinfo {volume} {02}},\ \bibinfo {pages} {093} (\bibinfo
  {year} {2007})},\ \Eprint {http://arxiv.org/abs/hep-ph/0611265}
  {arXiv:hep-ph/0611265} \BibitemShut {NoStop}%
\bibitem [{\citenamefont {Kopeliovich}\ \emph {et~al.}(2007)\citenamefont
  {Kopeliovich}, \citenamefont {Nemchik},\ and\ \citenamefont
  {Schmidt}}]{Kopeliovich:2006xy}%
  \BibitemOpen
  \bibfield  {author} {\bibinfo {author} {\bibfnamefont {B.}~\bibnamefont
  {Kopeliovich}}, \bibinfo {author} {\bibfnamefont {J.}~\bibnamefont
  {Nemchik}}, \ and\ \bibinfo {author} {\bibfnamefont {I.}~\bibnamefont
  {Schmidt}},\ }\href {\doibase 10.1016/j.nuclphysa.2006.10.059} {\bibfield
  {journal} {\bibinfo  {journal} {Nucl.Phys.}\ }\textbf {\bibinfo {volume}
  {A782}},\ \bibinfo {pages} {224} (\bibinfo {year} {2007})},\ \Eprint
  {http://arxiv.org/abs/hep-ph/0608044} {arXiv:hep-ph/0608044 [hep-ph]}
  \BibitemShut {NoStop}%
\bibitem [{\citenamefont {Armesto}\ \emph {et~al.}(2012)\citenamefont {Armesto}
  \emph {et~al.}}]{Armesto:2011ht}%
  \BibitemOpen
  \bibfield  {author} {\bibinfo {author} {\bibfnamefont {N.}~\bibnamefont
  {Armesto}} \emph {et~al.},\ }\href {\doibase 10.1103/PhysRevC.86.064904}
  {\bibfield  {journal} {\bibinfo  {journal} {Phys. Rev. C}\ }\textbf {\bibinfo
  {volume} {86}},\ \bibinfo {pages} {064904} (\bibinfo {year} {2012})},\
  \Eprint {http://arxiv.org/abs/1106.1106} {arXiv:1106.1106 [hep-ph]}
  \BibitemShut {NoStop}%
\bibitem [{\citenamefont {Arleo}(2002)}]{Arleo:2002kh}%
  \BibitemOpen
  \bibfield  {author} {\bibinfo {author} {\bibfnamefont {F.}~\bibnamefont
  {Arleo}},\ }\href {\doibase 10.1088/1126-6708/2002/11/044} {\bibfield
  {journal} {\bibinfo  {journal} {JHEP}\ }\textbf {\bibinfo {volume} {11}},\
  \bibinfo {pages} {044} (\bibinfo {year} {2002})},\ \Eprint
  {http://arxiv.org/abs/hep-ph/0210104} {arXiv:hep-ph/0210104} \BibitemShut
  {NoStop}%
\bibitem [{\citenamefont {Baier}\ \emph
  {et~al.}(1997{\natexlab{a}})\citenamefont {Baier}, \citenamefont
  {Dokshitzer}, \citenamefont {Mueller}, \citenamefont {Peigne},\ and\
  \citenamefont {Schiff}}]{Baier:1996kr}%
  \BibitemOpen
  \bibfield  {author} {\bibinfo {author} {\bibfnamefont {R.}~\bibnamefont
  {Baier}}, \bibinfo {author} {\bibfnamefont {Y.~L.}\ \bibnamefont
  {Dokshitzer}}, \bibinfo {author} {\bibfnamefont {A.~H.}\ \bibnamefont
  {Mueller}}, \bibinfo {author} {\bibfnamefont {S.}~\bibnamefont {Peigne}}, \
  and\ \bibinfo {author} {\bibfnamefont {D.}~\bibnamefont {Schiff}},\ }\href
  {\doibase 10.1016/S0550-3213(96)00553-6} {\bibfield  {journal} {\bibinfo
  {journal} {Nucl. Phys. B}\ }\textbf {\bibinfo {volume} {483}},\ \bibinfo
  {pages} {291} (\bibinfo {year} {1997}{\natexlab{a}})},\ \Eprint
  {http://arxiv.org/abs/hep-ph/9607355} {arXiv:hep-ph/9607355} \BibitemShut
  {NoStop}%
\bibitem [{\citenamefont {Baier}\ \emph {et~al.}(2001)\citenamefont {Baier},
  \citenamefont {Dokshitzer}, \citenamefont {Mueller},\ and\ \citenamefont
  {Schiff}}]{Baier:2001yt}%
  \BibitemOpen
  \bibfield  {author} {\bibinfo {author} {\bibfnamefont {R.}~\bibnamefont
  {Baier}}, \bibinfo {author} {\bibfnamefont {Y.~L.}\ \bibnamefont
  {Dokshitzer}}, \bibinfo {author} {\bibfnamefont {A.~H.}\ \bibnamefont
  {Mueller}}, \ and\ \bibinfo {author} {\bibfnamefont {D.}~\bibnamefont
  {Schiff}},\ }\href@noop {} {\bibfield  {journal} {\bibinfo  {journal} {JHEP}\
  }\textbf {\bibinfo {volume} {0109}},\ \bibinfo {pages} {033} (\bibinfo {year}
  {2001})},\ \Eprint {http://arxiv.org/abs/hep-ph/0106347}
  {arXiv:hep-ph/0106347 [hep-ph]} \BibitemShut {NoStop}%
\bibitem [{\citenamefont {Wang}\ \emph {et~al.}(1996)\citenamefont {Wang},
  \citenamefont {Huang},\ and\ \citenamefont {Sarcevic}}]{Wang:1996yh}%
  \BibitemOpen
  \bibfield  {author} {\bibinfo {author} {\bibfnamefont {X.-N.}\ \bibnamefont
  {Wang}}, \bibinfo {author} {\bibfnamefont {Z.}~\bibnamefont {Huang}}, \ and\
  \bibinfo {author} {\bibfnamefont {I.}~\bibnamefont {Sarcevic}},\ }\href
  {\doibase 10.1103/PhysRevLett.77.231} {\bibfield  {journal} {\bibinfo
  {journal} {Phys. Rev. Lett.}\ }\textbf {\bibinfo {volume} {77}},\ \bibinfo
  {pages} {231} (\bibinfo {year} {1996})},\ \Eprint
  {http://arxiv.org/abs/hep-ph/9605213} {arXiv:hep-ph/9605213} \BibitemShut
  {NoStop}%
\bibitem [{\citenamefont {Baier}\ \emph
  {et~al.}(1997{\natexlab{b}})\citenamefont {Baier}, \citenamefont
  {Dokshitzer}, \citenamefont {Mueller}, \citenamefont {Peigne},\ and\
  \citenamefont {Schiff}}]{Baier:1996sk}%
  \BibitemOpen
  \bibfield  {author} {\bibinfo {author} {\bibfnamefont {R.}~\bibnamefont
  {Baier}}, \bibinfo {author} {\bibfnamefont {Y.~L.}\ \bibnamefont
  {Dokshitzer}}, \bibinfo {author} {\bibfnamefont {A.~H.}\ \bibnamefont
  {Mueller}}, \bibinfo {author} {\bibfnamefont {S.}~\bibnamefont {Peigne}}, \
  and\ \bibinfo {author} {\bibfnamefont {D.}~\bibnamefont {Schiff}},\ }\href
  {\doibase 10.1016/S0550-3213(96)00581-0} {\bibfield  {journal} {\bibinfo
  {journal} {Nucl. Phys. B}\ }\textbf {\bibinfo {volume} {484}},\ \bibinfo
  {pages} {265} (\bibinfo {year} {1997}{\natexlab{b}})},\ \Eprint
  {http://arxiv.org/abs/hep-ph/9608322} {arXiv:hep-ph/9608322} \BibitemShut
  {NoStop}%
\bibitem [{\citenamefont {Baier}\ \emph {et~al.}(1998)\citenamefont {Baier},
  \citenamefont {Dokshitzer}, \citenamefont {Mueller},\ and\ \citenamefont
  {Schiff}}]{Baier:1998kq}%
  \BibitemOpen
  \bibfield  {author} {\bibinfo {author} {\bibfnamefont {R.}~\bibnamefont
  {Baier}}, \bibinfo {author} {\bibfnamefont {Y.~L.}\ \bibnamefont
  {Dokshitzer}}, \bibinfo {author} {\bibfnamefont {A.~H.}\ \bibnamefont
  {Mueller}}, \ and\ \bibinfo {author} {\bibfnamefont {D.}~\bibnamefont
  {Schiff}},\ }\href {\doibase 10.1016/S0550-3213(98)00546-X} {\bibfield
  {journal} {\bibinfo  {journal} {Nucl.Phys.}\ }\textbf {\bibinfo {volume}
  {B531}},\ \bibinfo {pages} {403} (\bibinfo {year} {1998})},\ \Eprint
  {http://arxiv.org/abs/hep-ph/9804212} {arXiv:hep-ph/9804212 [hep-ph]}
  \BibitemShut {NoStop}%
\bibitem [{\citenamefont {Ciofi~degli Atti}\ and\ \citenamefont
  {Simula}(1996)}]{CiofidegliAtti:1995qe}%
  \BibitemOpen
  \bibfield  {author} {\bibinfo {author} {\bibfnamefont {C.}~\bibnamefont
  {Ciofi~degli Atti}}\ and\ \bibinfo {author} {\bibfnamefont {S.}~\bibnamefont
  {Simula}},\ }\href {\doibase 10.1103/PhysRevC.53.1689} {\bibfield  {journal}
  {\bibinfo  {journal} {Phys. Rev. C}\ }\textbf {\bibinfo {volume} {53}},\
  \bibinfo {pages} {1689} (\bibinfo {year} {1996})},\ \Eprint
  {http://arxiv.org/abs/nucl-th/9507024} {arXiv:nucl-th/9507024} \BibitemShut
  {NoStop}%
\bibitem [{\citenamefont {Kovarik}\ \emph {et~al.}(2016)\citenamefont {Kovarik}
  \emph {et~al.}}]{Kovarik:2015cma}%
  \BibitemOpen
  \bibfield  {author} {\bibinfo {author} {\bibfnamefont {K.}~\bibnamefont
  {Kovarik}} \emph {et~al.},\ }\href {\doibase 10.1103/PhysRevD.93.085037}
  {\bibfield  {journal} {\bibinfo  {journal} {Phys. Rev. D}\ }\textbf {\bibinfo
  {volume} {93}},\ \bibinfo {pages} {085037} (\bibinfo {year} {2016})},\
  \Eprint {http://arxiv.org/abs/1509.00792} {arXiv:1509.00792 [hep-ph]}
  \BibitemShut {NoStop}%
\bibitem [{\citenamefont {de~Florian}\ \emph {et~al.}(2015)\citenamefont
  {de~Florian}, \citenamefont {Sassot}, \citenamefont {Epele}, \citenamefont
  {Hern\'andez-Pinto},\ and\ \citenamefont {Stratmann}}]{deFlorian:2014xna}%
  \BibitemOpen
  \bibfield  {author} {\bibinfo {author} {\bibfnamefont {D.}~\bibnamefont
  {de~Florian}}, \bibinfo {author} {\bibfnamefont {R.}~\bibnamefont {Sassot}},
  \bibinfo {author} {\bibfnamefont {M.}~\bibnamefont {Epele}}, \bibinfo
  {author} {\bibfnamefont {R.~J.}\ \bibnamefont {Hern\'andez-Pinto}}, \ and\
  \bibinfo {author} {\bibfnamefont {M.}~\bibnamefont {Stratmann}},\ }\href
  {\doibase 10.1103/PhysRevD.91.014035} {\bibfield  {journal} {\bibinfo
  {journal} {Phys. Rev. D}\ }\textbf {\bibinfo {volume} {91}},\ \bibinfo
  {pages} {014035} (\bibinfo {year} {2015})},\ \Eprint
  {http://arxiv.org/abs/1410.6027} {arXiv:1410.6027 [hep-ph]} \BibitemShut
  {NoStop}%
\bibitem [{\citenamefont {Boglione}\ \emph {et~al.}(2022)\citenamefont
  {Boglione}, \citenamefont {Diefenthaler}, \citenamefont {Dolan},
  \citenamefont {Gamberg}, \citenamefont {Melnitchouk}, \citenamefont
  {Pitonyak}, \citenamefont {Prokudin}, \citenamefont {Sato},\ and\
  \citenamefont {Scalyer}}]{Boglione:2022gpv}%
  \BibitemOpen
  \bibfield  {author} {\bibinfo {author} {\bibfnamefont {M.}~\bibnamefont
  {Boglione}}, \bibinfo {author} {\bibfnamefont {M.}~\bibnamefont
  {Diefenthaler}}, \bibinfo {author} {\bibfnamefont {S.}~\bibnamefont {Dolan}},
  \bibinfo {author} {\bibfnamefont {L.}~\bibnamefont {Gamberg}}, \bibinfo
  {author} {\bibfnamefont {W.}~\bibnamefont {Melnitchouk}}, \bibinfo {author}
  {\bibfnamefont {D.}~\bibnamefont {Pitonyak}}, \bibinfo {author}
  {\bibfnamefont {A.}~\bibnamefont {Prokudin}}, \bibinfo {author}
  {\bibfnamefont {N.}~\bibnamefont {Sato}}, \ and\ \bibinfo {author}
  {\bibfnamefont {Z.}~\bibnamefont {Scalyer}} (\bibinfo {collaboration}
  {Jefferson Lab Angular Momentum (JAM)}),\ }\href {\doibase
  10.1007/JHEP04(2022)084} {\bibfield  {journal} {\bibinfo  {journal} {JHEP}\
  }\textbf {\bibinfo {volume} {04}},\ \bibinfo {pages} {084} (\bibinfo {year}
  {2022})},\ \Eprint {http://arxiv.org/abs/2201.12197} {arXiv:2201.12197
  [hep-ph]} \BibitemShut {NoStop}%
\bibitem [{\citenamefont {Airapetian}\ \emph
  {et~al.}(2013{\natexlab{a}})\citenamefont {Airapetian} \emph
  {et~al.}}]{HERMES:2012kpt}%
  \BibitemOpen
  \bibfield  {author} {\bibinfo {author} {\bibfnamefont {A.}~\bibnamefont
  {Airapetian}} \emph {et~al.} (\bibinfo {collaboration} {HERMES}),\ }\href
  {\doibase 10.1103/PhysRevD.87.012010} {\bibfield  {journal} {\bibinfo
  {journal} {Phys. Rev. D}\ }\textbf {\bibinfo {volume} {87}},\ \bibinfo
  {pages} {012010} (\bibinfo {year} {2013}{\natexlab{a}})},\ \Eprint
  {http://arxiv.org/abs/1204.4161} {arXiv:1204.4161 [hep-ex]} \BibitemShut
  {NoStop}%
\bibitem [{\citenamefont {Airapetian}\ \emph
  {et~al.}(2013{\natexlab{b}})\citenamefont {Airapetian} \emph
  {et~al.}}]{HERMES:2012uyd}%
  \BibitemOpen
  \bibfield  {author} {\bibinfo {author} {\bibfnamefont {A.}~\bibnamefont
  {Airapetian}} \emph {et~al.} (\bibinfo {collaboration} {HERMES}),\ }\href
  {\doibase 10.1103/PhysRevD.87.074029} {\bibfield  {journal} {\bibinfo
  {journal} {Phys. Rev. D}\ }\textbf {\bibinfo {volume} {87}},\ \bibinfo
  {pages} {074029} (\bibinfo {year} {2013}{\natexlab{b}})},\ \Eprint
  {http://arxiv.org/abs/1212.5407} {arXiv:1212.5407 [hep-ex]} \BibitemShut
  {NoStop}%
\bibitem [{\citenamefont {James}\ and\ \citenamefont
  {Roos}(1975)}]{James:1975dr}%
  \BibitemOpen
  \bibfield  {author} {\bibinfo {author} {\bibfnamefont {F.}~\bibnamefont
  {James}}\ and\ \bibinfo {author} {\bibfnamefont {M.}~\bibnamefont {Roos}},\
  }\href {\doibase 10.1016/0010-4655(75)90039-9} {\bibfield  {journal}
  {\bibinfo  {journal} {Comput. Phys. Commun.}\ }\textbf {\bibinfo {volume}
  {10}},\ \bibinfo {pages} {343} (\bibinfo {year} {1975})}\BibitemShut
  {NoStop}%
\end{thebibliography}%

\end{document}